\documentclass[12pt,draftclsnofoot,letterpaper,onecolumn,romanappendices]{IEEEtran}
\usepackage{cite}
\usepackage{epstopdf}
\usepackage{epsfig}
\usepackage{amsmath,amssymb,amsfonts,amstext,amsbsy,amsopn,dsfont}
\usepackage{cases}
\usepackage{sublabel}
\usepackage{color}
\usepackage{array}

\newtheorem{theorem}{\textbf{Theorem}}

\newtheorem{lemma}{\textbf{Lemma}}

\newtheorem{definition}{\textbf{Definition}}
\newtheorem{corollary}{\textbf{Corollary}}

\newcommand{\defn}{\triangleq}
\newcommand{\dif}{\textmd{d}}
\newcommand{\Lt}{\mathsf{L}}
\newcommand{\Ut}{\mathsf{U}}
\newcommand{\Rt}{\mathsf{R}}

\newcommand{\cov}{\mathrm{coc}}

\begin{document}

\title{On the Limits of Coexisting Coverage and Capacity in Multi-RAT Heterogeneous Networks}

\author{Chun-Hung Liu and Hong-Cheng Tsai\\
\thanks{C.-H. Liu and H.-C. Tsai are with the Department of Electrical and Computer Engineering at National Chiao Tung University, Hsinchu, Taiwan. The contact author is Prof. Liu  (e-mail: chungliu@nctu.edu.tw). Part of this work was presented in IEEE International Conference on Communications (ICC), May 2016 \cite{HCTCHL16}. Revised Manuscript Date: \today.}
}

\maketitle

\begin{abstract}
This paper devises a general modeling and analyzing framework for  a  heterogeneous wireless network (HetNet) in which several wireless subnetworks coexist and use multiple radio access technologies (multi-RATs). The coexisting coverage and network capacity in such a multi-RAT HetNet are hardly investigated in prior works. To characterize the coexisting interactions in a multi-RAT HetNet, in this paper we consider a HetNet consisting of $K$ tiers of APs and two different RATs, RAT-$\Lt$ and RAT-$\Ut$, are adopted in the HetNet. RAT-$\Lt$ is adopted by the access points (APs) in the first $K-1$ tiers and APs in the $K$th tier only use RAT-$\Ut$. Both noncrossing-RAT and crossing-RAT user association scenarios are considered. In each scenario, the void probability and channel access probability of the APs in each tier are first found and then the tight lower bounds and their lowest limits on the proposed coexisting coverage and network capacity are derived. We show that multi-RAT networks in general can achieve higher link coverage and capacity by using opportunistic CSMA/CA that avoids/alleviates severe interfering between all coexisting APs. Also, crossing-RAT user association is shown to achieve much higher coexisting coverage and network capacity than noncrossing-RAT user association. Finally, numerical simulations for the LTE-U and WiFi networks coexisting in the HetNet validate our findings.

\end{abstract}

\section{Introduction}

Cellular networks consisting of several different kinds of macro and small cell base stations (BSs) have increasingly become a prevailing network topology that is able to remarkably enhance network coverage and throughput due to the dense deployment of small cell BSs. Such a heterogeneous cellular network, sooner or later, will be seamlessly incorporated with other wireless networks of different radio access technologies (RATs) in oder to comprehensively improve wireless services from different aspects of demands. For example, the Internet of Things (IoT) has to integrate many different wireless networks using different kinds of wireless transmission technologies and interfaces in order to remotely control different physical objects and devices\cite{AAWHMU15}. To fulfill such a large-scale networking system like IoT, we need to understand the fundamental properties and limits of \textit{heterogeneous} wireless networks in which there coexist different kinds of wireless access points (APs) that adopt multiple distinct RATs. Therefore, how to tractably model and analyze this kind of heterogeneous wireless networks is an interesting and important problem worth investigating thoroughly. Recently, modeling a heterogeneous cellular network by using the stochastic geometry framework has made a great stride on the tractable analysis of the transmission performance metrics, such as signal-to-interference ratio (SIR), link coverage, throughput and energy efficiency (typically see \cite{HSDRKGFBJGA12,HSJYJSXPJGA12,PXCHLJGA13,CHL14,CHLKLF6}). Almost all the prior works on heterogeneous wireless networks primarily focus on the single-RAT modeling and analysis so that their findings cannot be extended and applied in a multi-RAT scenario straightway. For example, channel access protocols for different RATs may be fairly distinct in principle and thus the transmission interactions between the APs using multiple RATs and distinct channel access protocols are definitely unable to be completely characterized by a single-RAT modeling framework. 

To tractably study the fundamental transmission performance in a multi-RAT heterogeneous wireless network (HetNet), in this paper we propose a very general approach to modeling and analyzing a multi-RAT HetNet consisting of $K$-tier APs. The APs in each tier are of the same type and performance, and they form an independent homogeneous Poisson point process (PPP) with a certain intensity. Two RATs, i.e., RAT-$\Lt$ and RAT-$\Ut$, operating on  non-overlapped frequency bands are adopted in this network\footnote{This multi-RAT network modeling can be referred to a practical case: the RAT-$\Lt$ APs can be the macro/small cell base stations using the licensed frequency band, whereas the RAT-$\Ut$ APs  can be the WiFi APs using the unlicensed frequency band.}, and specifically all the APs in the first $K-1$ tiers primarily adopt RAT-$\Lt$ and opportunistically use RAT-$\Ut$  and the APs in the $K$th tier only use RAT-$\Ut$. All APs in the first $K-1$ tiers can access the RAT-$\Lt$ channel at will without contending whereas all the APs in the networks have to contend the RAT-$\Ut$ channel before accessing it by using the (slotted nonpersistant) opportunistic CSMA/CA with random backoff time protocol. Such a general multi-RAT network model that can characterize the coexisting scenario of multiple large-scale random networks using different RATs as well as channel access protocols has not yet been studied in the literature.

\subsection{Motivation and Prior Work}
A few earlier  prior works on investigating the coexistence issue in multiple wireless networks mainly focused on how to efficiently and fairly share the unlicensed bands. In \cite{REAPDT07}, a game-theoretical approach was proposed to solve the spectrum sharing problem for multiple coexisting and interfering networks. References  \cite{HYPPHCNRP07,YMSKAH09,JBENNJR12} characterized the interference modeling and mitigation in the unlicensed bands. These works are not developed in a large-scale network model and the fundamental coexisting issues, such as the success transmission problem and network throughput, are not studied. A more accurate interference analysis technique based on the continuum field approximation and spiral representation was proposed in \cite{JJQLHNAPGW14} for large-scale networks, but it still does not characterize the fundamental relationship between the interference and the intensities (densities) of the wireless APs using different RATs.

Recently, the coexistence problem in multi-RAT wireless networks has been gained more and more attentions since studying this problem helps different RAT networks jointly improve their wireless resource utilization. For example, LTE and WiFi networks can coexist in the unlicensed band to significantly improve the network capacity\cite{HZXCWGSW15,RZMWLXCZZXSLLX15}. To alleviate the coexisting interference impact in these two different kinds of wireless networks, the effective approach is either to offload traffic from LTE to WiFi networks or to make these two systems share the unlicensed spectrum resource in an appropriate way (see a recent work in \cite{QCGYHSAMGYLAH16} for this study). Purely offloading traffic from an LTE network to another WiFi network could not effectively improve the total capacity of these two networks when the WiFi network has limited resource for external offloading. On the contrary, if these two networks can coexist without causing severe interference, their sum capacity can be significantly improved.  Reference \cite{EARCWPIADZ13} showed that small cell BSs have a notable throughput gain if they can adaptively access the unlicensed band without affecting the WiFi APs. An adaptive channel access protocol based on listen-before-talk (LBT) for coexisting LTE-U and WiFi networks was proposed in \cite{RYGYAMGYL16}. It can adaptively adjust the backoff window size according to the available licensed spectrum bandwidth so that the network throughput is improved. However, such a protocol may not effectively improve the throughput of a large-scale dense network since it does not exclude the APs with bad channel conditions that occupy the unlicensed spectrum resource. In \cite{ABCIPZ14,SSISDR15,XDCHLLCWXZ16,CHLLCW1602}, stochastic geometry is applied to analyze the coexistence performance of large-scale LTE and WiFi networks, but the network models in these works are too simple to completely characterize the discrepancies originating from different RATs, such as distinct channel access protocols, different user association schemes for different RATs, etc. Hence, the analytical results in these works may be far away from their corresponding realistic outcomes.

\subsection{Contributions}
To study the fundamental limits on the coexistence performance in multi-RAT HetNets, in this paper our first contribution is to propose a very general model for a large-scale heterogeneous wireless network where $K$ different types of APs that independently form multiple overlaid homogeneous PPPs adopt two distinct channel access protocols for RAT-$\Ut$ and RAT-$\Lt$. To thoroughly evaluate the coexisting transmission performance in the multi-RAT HetNet, specifically we consider a generalized user association scheme for such a multi-RAT HeNet under noncrossing-RAT and crossing-RAT user association scenarios. Our second important contribution is to first derive the accurate void probability of the APs in each tier under the two considered user association scenarios and to show that the void probabilities depending on the user and AP intensities are no longer negligible in a densely deployed network, which has a significant impact on the interference modeling and cell load analysis. Then the exact channel access probability for the opportunistic CSMA/CA with random backoff time protocol is found for each association scenario, and it is so general that it can be used to calculate all CSMA/CA-based channel access probabilities. 

In the noncrossing-RAT scenario, users cannot associate with an AP that primarily uses the RAT different from the RAT they adopt\footnote{This scenario can be exemplified by a practical situation that LTE users cannot associate with an WiFi AP, or WiFi users cannot associate with an LTE BS.}. On the contrary,  in the crossing-RAT scenario users can associate with an AP no matter which RAT the AP adopts. For each user association scenario, the link coverage (probability) of the APs in each tier using the RAT-$\Lt$ channel or the RAT-$\Ut$ channel or both is theoretically shown to be very close to its closed-form lower bound derived by assuming void APs (i.e., APs are not tagged by any users) still can be described by thinning independent PPPs. Also, as the user intensity goes to infinity the link coverage is shown to reduce and converge to a constant that is its fundamental lowest limit, which indicates the link  coverage would be significantly underestimated provided that the cell voidness issue is not considered in the model. This is our third contribution.

Our final contribution is to propose the coexisting coverage and network capacity metrics defined based on the link coverages and mean spectrum efficiencies of the APs using the two RATs in different tiers. According to the tight lower bounds on the link coverages, the tight lower bounds on the coexisting coverage and network capacity can be easily derived, which provide the overall average network coverage and capacity that can give us some insights into how to design channel access protocols for different RATs and how densely to deploy APs while they are using different RATs. A numerical simulation example is given for applying our modeling and analysis framework to the HetNet where LTE small cell BSs and WiFi APs coexist. It not only validates the correctness and accurateness of all derived coverage and capacity results, but also importantly indicates how much network-wise capacity gain can be exploited for coexisting LTE-U BSs and WiFi APs. 

\section{Network Model and Preliminaries}
\subsection{Multi-RAT Heterogeneous Network Modeling}
Suppose a large-scale planar heterogeneous wireless network consisting of  $K$ tiers of access points (APs, or called base stations)\footnote{The concept of a tier of APs here means the same type of APs consisting of one tier. Thus, the entire heterogeneous network consists of $K$ different types of APs.}. To reduce the analysis of complexity, only two radio access technologies (RATs), RAT-$\Lt$ and RAT-$\Ut$, are adopted in the network, and they are operated in two different non-overlapped frequency bands. The first $K-1$ tiers in the network consist of the APs \textbf{\textit{primarily}} \textit{adopting RAT-$\Lt$} and \textbf{\textit{opportunistically}} \textit{adopting RAT-$\Ut$ }if they are able to access the RAT-$\Ut$ channel\footnote{Please be aware that here the RAT-$\Lt$ APs are said to opportunistically adopt RAT-$\Ut$ since their users associate with them by using their channel state information in the RAT-$\Lt$ frequency band only.}, and the $K$th tier consists of the APs only adopting RAT-$\Ut$  in the network. Specifically, all APs in the $k$th tier form a marked homogeneous Poisson point process (PPP) of intensity $\lambda_k$ denoted by 
\begin{align}
\Phi_k \defn\{&(X_{k,i},P_k,V_{k,i}): X_{k,i}\in\mathbb{R}^2, P_k\in\mathbb{R}_+, V_{k,i}\in\{0,1\}, \forall i\in\mathbb{N}_+, k\in\mathcal{K}\defn \{1,\ldots, K\}\},
\end{align}
where $X_{k,i}$ denotes AP $i$ in the $k$th tier and its location, $P_k$ is the transmit power of all tier-$k$ APs, $V_{k,i}$ is a Bernoulli random variable that is zero if AP $X_{k,i}$ is void and one otherwise\footnote{The distribution of $V_k$ is affected by the user association mentioned in the following and its closed-form expression is given in Lemma \ref{Lem:VoidProb}.}. Table \ref{Tab:MathNotation} summarizes the notations of all main variables and functions used in this paper.  

\begin{table}[!t]
\centering
	\caption{Notation of Main Variables and Functions}\label{Tab:MathNotation}
	\begin{tabular}{|c|c|}
		\hline
		Symbol & Meaning\\ \hline
		$\Phi_k$ &  Homogeneous PPP of the tier-$k$ APs\\
		$\Phi$ &  $\bigcup_{k=1}^K \Phi_k$\\
	    $X_{k,i}$ &  AP $i$ in the $k$th tier and its location\\
	    $P_k$ &  Transmit power of the tier-$k$ APs\\
		$\lambda_k$ & Intensity of the tier-$k$ APs\\
		$H_{k,i}$ & Rayleigh fading channel gain of AP $X_{k,i}$\\
		$G^{-1}_{k,i}$ & Log-normal shadowing gain of AP $X_{k,i}$\\
		$V_{k,i}\in\{0,1\}$ & One if AP $X_{k,i}$ not void and zero otherwise\\
		$W_{k,i}$ & (Random) association weight of AP $X_{k,i}$\\
		$\alpha>2$ & Pathloss exponent\\
		$\mu_{\Rt}$  & Intensity of the RAT-$\Rt$ users (noncrossing-RAT), $\Rt\in\{\Ut,\Lt\}$ \\
		$\mu$ & Intensity of total users ($\mu=\mu_{\Rt}+\mu_{\Lt}$, crossing-RAT)\\
		$\nu_k (\hat{\nu}_k)$  & Void probability of tier  $k$ for noncrossing (crossing)-RAT\\
		$\zeta_k$ & $\frac{7}{2}\mathbb{E}\left[W^{\frac{2}{\alpha}}_k\right]\mathbb{E}\left[W^{\frac{2}{\alpha}}_k\right]$\\
		$\vartheta_k$  & $\lambda_k\mathbb{E}\left[W^{\frac{2}{\alpha}}_k\right]/\sum_{m=1}^{K-1}\lambda_m\mathbb{E}\left[W^{\frac{2}{\alpha}}_m\right]$  (noncrossing-RAT)\\
		$\hat{\vartheta}_k$  &
		$\lambda_k\mathbb{E}\left[W^{\frac{2}{\alpha}}_k\right]/\sum_{m=1}^{K}\lambda_m\mathbb{E}\left[W^{\frac{2}{\alpha}}_m\right]$  (crossing-RAT)\\
		$\mathcal{S}_k$ & (Random) sensing region of the tier-$k$ APs\\
		$\delta(\mathcal{A})$ & Lebesgue measure (area) of set $\mathcal{A}$\\
		$\Delta$ & Channel gain threshold for Opportunistic CSMA/CA\\
		$p_k$ & $\mathbb{P}[H_kG^{-1}_k\geq \Delta]$\\
		$T_k\in[0,\tau_k]$ & Random backoff time duration of the tier-$k$ APs\\
		$\rho_k (\hat{\rho}_k)$ & Channel access probability of tier $k$, noncrossing (crossing)-RAT\\
		$\theta$ & SIR threshold for Coverage  \\
		$P_{\Rt} (\hat{P}_{\Rt})$ & Coverage of RAT-$\Rt$, noncrossing (crossing)-RAT  \\
		$\mathsf{C}_{\Rt} (\hat{\mathsf{C}}_{\Rt})$ & Ergodic rate of RAT-$\Rt$, noncrossing (crossing)-RAT\\
		$f_Z(\cdot) (F_Z(\cdot))$ & pdf (CDF) of random variable $Z$\\
		\hline
	\end{tabular}
\end{table}

In this paper, we consider two distinct scenarios of user association: noncrossing-RAT user association and crossing-RAT user association.  In the scenario of noncrossing-RAT user association, there are two kinds of RAT-$\Lt$ and RAT-$\Ut$ users in the network: the RAT-$\Lt$ users who are assumed to form an independent homogeneous PPP of intensity $\mu_{\Lt}$ can only associate with the APs in the first $K-1$ tiers, whereas the RAT-$\Ut$ users who form another independent homogeneous PPP of intensity $\mu_{\Ut}$ only associate with the tier-$K$ APs. Under this scenario, the user association scheme based on an RAT-$\Rt$ typical user located at the origin can be written as
\begin{align}\label{Eqn:NonCrossUserAssSch}
X_{\Rt}^*\defn \begin{cases}
X^*_{\Lt}\defn \arg\sup_{k,i:X_{k,i}\in\Phi\setminus\Phi_K} W_{k,i}\|X_{k,i}\|^{-\alpha}, \text{if }\Rt=\Lt\\
X^*_{\Ut}\defn\arg\sup_{K,i:X_{K,i}\in\Phi_K} W_{K,i}\|X_{K,i}\|^{-\alpha}, \text{if }\Rt=\Ut
\end{cases}
\end{align}
in which $\Rt\in\{\Lt,\Ut\}$, $X^*_{\Rt}$ stands for the associated RAT-$\Rt$ AP and its location, $\Phi\defn \bigcup^K_{k=1}\Phi_k$, $\|X-Y\|$ denotes the Euclidean distance between nodes $X$ and $Y$, $W_{k,i}$ is called the (random) association weight of AP $X_{k,i}$\footnote{If $W_{k,i}$  is related to the channel gain of AP $X_{k,i}$ primarily adopting RAT-$\Lt$, it must be only related to the RAT-$\Lt$ channel gain of $X_{k,i}$ even when $X_{k,i}$ can opportunistically access the RAT-$\Ut$ channel.}, and $\alpha>2$ is the path loss exponent\footnote{To simplify our following analysis, we assume that the pathloss exponents in the RAT-$\Lt$ and the RAT $\Ut$ frequency bands are the same $\alpha$, whereas such an assumption is commonly used in the simulation setting of licensed and unlicensed bands, for example, see the technical report of the LTE-U forum in \cite{LTEU-Report}.}. The distance from $X^*_{\Rt}$ to the origin can be expressed as 
\begin{align}\label{Eqn:NonCrossUserAssSchDist}
\|X_{\Rt}^*\|= \begin{cases}
\|X^*_{\Lt}\|\defn (W^*_{\Lt})^{\frac{1}{\alpha}}\left(\inf_{k,i:X_{k,i}\in\Phi\setminus\Phi_K} W^{-\frac{1}{\alpha}}_{k,i}\|X_{k,i}\|\right), \text{if }\Rt=\Lt\\
\|X^*_{\Ut}\|\defn (W^*_{\Ut})^{\frac{1}{\alpha}}\left(\inf_{K,i:X_{K,i}\in\Phi_K} W^{-\frac{1}{\alpha}}_{K,i}\|X_{K,i}\|\right), \text{if }\Rt=\Ut
\end{cases},
\end{align}
where $W^*_{\Rt}$ is the association weight of $X^*_{\Rt}$. 

For the scenario of crossing-RAT user association, all users are assumed to form an independent homogeneous PPP of intensity $\mu$ and they can associate with any one of the RAT-$\Ut$ and RAT-$\Lt$ APs by the following association scheme based on a typical user located at the origin:
\begin{align}\label{Eqn:CrossUserAss}
X^*\defn\arg\sup_{k,i:X_{k,i}\in\Phi}W_{k,i}\|X_{k,i}\|^{-\alpha}=\arg\inf_{k,i:X_{k,i}\in\Phi}W^{-\frac{1}{\alpha}}_{k,i}\|X_{k,i}\|,
\end{align}
where $X^*$ denotes the associated AP and its location. The distance from $X^*$ to the typical user located at the origin can be equivalently written as
\begin{align}\label{Eqn:CrossUserAssDis}
\|X^*\|=(W^*)^{\frac{1}{\alpha}}\left(\inf_{k,i:X_{k,i}\in\Phi}W^{-\frac{1}{\alpha}}_{k,i}\|X_{k,i}\|\right),
\end{align}
where $W^*\in\{W_{k,i}, \forall i\in\mathbb{N}_+, k\in\mathcal{K}\}$ is the association weight of $X^*$. The user association schemes in \eqref{Eqn:NonCrossUserAssSch} and \eqref{Eqn:CrossUserAss} both are a weighted power-law design of pathloss. They are so general that they are able to cover several user association schemes \cite{CHLLCW16,CHLLCW1502}. For example, we can have the biased nearest AP association (BNA) scheme if all $W_{k,i}$'s are constant, like the prior works in \cite{HSDRKGFBJGA12}\cite{HSJYJSXPJGA12}. Or letting $W_{k,i}=P_kG^{-1}_{k,i}$ yields the mean maximum received-power association  (MMPA) scheme by assuming that APs only average out small-scale fading and still leave large-scale shadowing gain $G^{-1}_{k,i}$ in the channel gain\footnote{For example, APs could not estimate the mean received
powers from non-stationary users that are moving very fast\cite{STUBER01}.}. Our following theoretical analyses on link coverage (probability) and ergodic link capacity are based on the MMPA scheme since it is a more implementable one in practice that exploits shadowing effects to increase the signal-to-interference ratio (SIR) and benefits the fundamental analyses of the limits on the link coverage and ergodic link capacity at users. 

The user association schemes in \eqref{Eqn:NonCrossUserAssSch} and \eqref{Eqn:CrossUserAss} are essentially user-centric so that they cannot ensure that every AP in the HetNet is always associated with at least one user, i.e., some APs may be \textit{void}\cite{CHLLCW1502,CHLLCW16}. Identifying whether an AP is void or not is very important for the SIR analysis since a void AP actually does not generate any interference. The void probability of the APs in the network is shown in the following lemma.
\begin{lemma}\label{Lem:VoidProb}
	Consider the scenario of noncrossing-RAT user association. The void probability of a tier-$k$ AP, i.e., $\nu_k=\mathbb{P}[V_{k,i}=0]$, is accurately shown as
	\begin{align}\label{Eqn:VoidProbNonCrossRAT}
	\nu_k=\left(1+\frac{\mu_{\Lt}\vartheta_k}{\zeta_k\lambda_k}\right)^{-\zeta_k}\mathds{1}(k\neq K)+\left(1+\frac{\mu_{\Ut}}{\zeta_K \lambda_K }\right)^{-\zeta_K}\mathds{1}(k=K),\,\, \forall k\in\mathcal{K},
	\end{align}
	where $\mathds{1}(\mathcal{E})$ is the indicator function that is equal to one if event $\mathcal{E}$ is true and zero otherwise, $\zeta_k\defn \frac{7}{2}\mathbb{E}\left[W^{\frac{2}{\alpha}}_k\right]\mathbb{E}\left[W^{-\frac{2}{\alpha}}_k\right]$, and $\vartheta_k=\lambda_k\mathbb{E}\left[W^{\frac{2}{\alpha}}_k\right]/\sum_{m=1}^{K-1}\lambda_m\mathbb{E}\left[W^{\frac{2}{\alpha}}_m\right]$ is the probability that a user associates with a tier-$k$ AP. For the scenario of crossing-RAT user association, the void probability of a tier-$k$ AP is accurately expressed as
	\begin{align}\label{Eqn:VoidProbCrossRAT}
	\hat{\nu}_k=\left(1+\frac{\mu\hat{\vartheta}_k}{\zeta_k\lambda_k}\right)^{-\zeta_k},\,\, \forall k\in\mathcal{K},
	\end{align}  
	where $\hat{\vartheta}_k=\lambda_k\mathbb{E}\left[W^{\frac{2}{\alpha}}_k\right]/\sum_{m=1}^{K}\lambda_m\mathbb{E}\left[W^{\frac{2}{\alpha}}_m\right]$.
\end{lemma} 
\begin{IEEEproof}
First consider the voidness issue of the APs in the $K$th tier with noncrossing-RAT user association. Since the RAT-$\Ut$ users only associate with the tier-$K$ APs, the void probability of the tier-$K$ APs is essentially a single-tier void probability problem. Such a single-tier void probability has been found and shown in Proposition 1 of our previous work in [20]. Thus, the void probability of the tier-$K$ APs can be readily obtained based on the result in Proposition 1 in \cite{CHLLCW16}, i.e., $\nu_K=\left(1+\mu_{\Ut}/\zeta_K \lambda_K \right)^{-\zeta_K}$, which is characterized by the normalized cell load $\mu_{\Ut}/\zeta_K\lambda_K$ that is the average number of the RAT-$\Ut$ users associated with a tier-$K$ AP normalized by $\zeta_K$. Motivated by the void probability of a tier-$K$ AP, for the APs in the first $K-1$ tiers, we also can characterize the void probability of a tier-$k$ AP by the average cell load of the tier-$k$ AP normalized by $\zeta_k$, which is $\mu_{\Lt}\vartheta_k/\lambda_k$, because $\mu_{\Lt}\vartheta_k/\lambda_k$ represents the fraction of the RAT-$\Lt$ users associated with the tier-$k$ APs. Hence, for $k\neq K$, the void probability of a tier-$k$ AP can be inferred as $\nu_k=(1+\mu_{\Lt}\vartheta_k/\zeta_k\lambda_k)^{-\zeta_k}$. For the scenario of crossing-RAT user association, the void probability in \eqref{Eqn:VoidProbNonCrossRAT} reduces to \eqref{Eqn:VoidProbCrossRAT} since all users can associate with an AP in any tier of the network no mater which RAT is primarily adopted by the AP.
\end{IEEEproof}

\begin{figure}[!t]
\centering
\includegraphics[width=4.5in,height=2.75in]{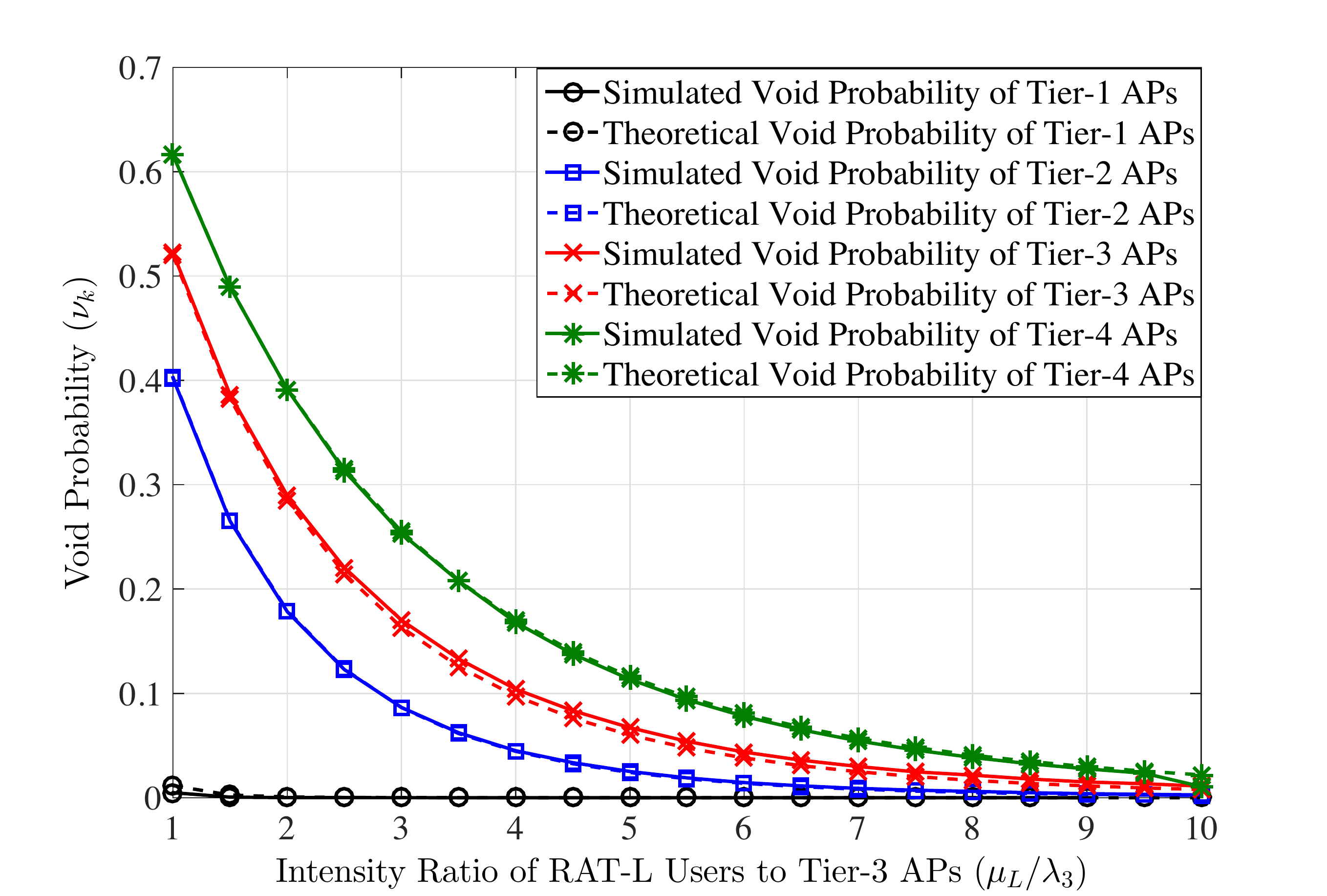}
\caption{Simulation results of the void probabilities of the APs in a four-tier multi-RAT HetNet using the MMPA scheme under the scenario of noncrossing-RAT user association. The network parameters for simulation are $\lambda_1=1.0\times 10^{-6}\text{ APs/m}^2$, $\lambda_1:\lambda_2:\lambda_3:\lambda_4=1:10:50:100$, $P_1=40$W, $P_2=1$W, $P_3=0.5$W, $P_4=0.2$W, $\alpha=4$, $\mu_{\Lt}=\mu_{\Ut}$, and $W_{k,i}=P_kG^{-1}_{k,i}$ where shadowing gain $G_{k,i}\sim\ln\mathcal{N}(0,\mathrm{3dB})$ is a log-normal random variable with  mean zero and variance $3$dB.}
\label{Fig:VoidProbHetNet}
\end{figure}

To validate the accurateness of the results in \eqref{Eqn:VoidProbNonCrossRAT}, the simulation results of the void probabilities of the APs in a four-tier HetNet are shown in Fig. \ref{Fig:VoidProbHetNet} and the network parameters for simulation are specified in the caption of Fig. \ref{Fig:VoidProbHetNet}. As shown in the figure, the void probability in \eqref{Eqn:VoidProbNonCrossRAT} is very accurate since it perfectly coincides with the simulation results. Most importantly, the figure illustrates that the void probabilities of the APs in the last three tiers are not small at all when the intensities of the APs in last three tiers are close to the user intensity. Thus, \textit{the void cell phenomenon should be considered while modeling the network performance metrics pertaining to the interference in a densely deployed network}. In addition to the aforementioned user association schemes and their induced void AP issue, another key point that needs to be specified is how the APs access their channels of the two RATs since channel access protocols dominate the interference modeling results. In the following subsection, an approach to modeling inconsistent random channel access based on the protocol of \textit{opportunistic} carrier sense multiple access with collision avoidance (CSMA/CA) will be introduced for the multi-RAT HetNet. 

\subsection{Channel Access Protocols for the Multi-RAT HetNet}

\begin{figure}[!t]
	\centering
	\includegraphics[width=4in,height=2.0in]{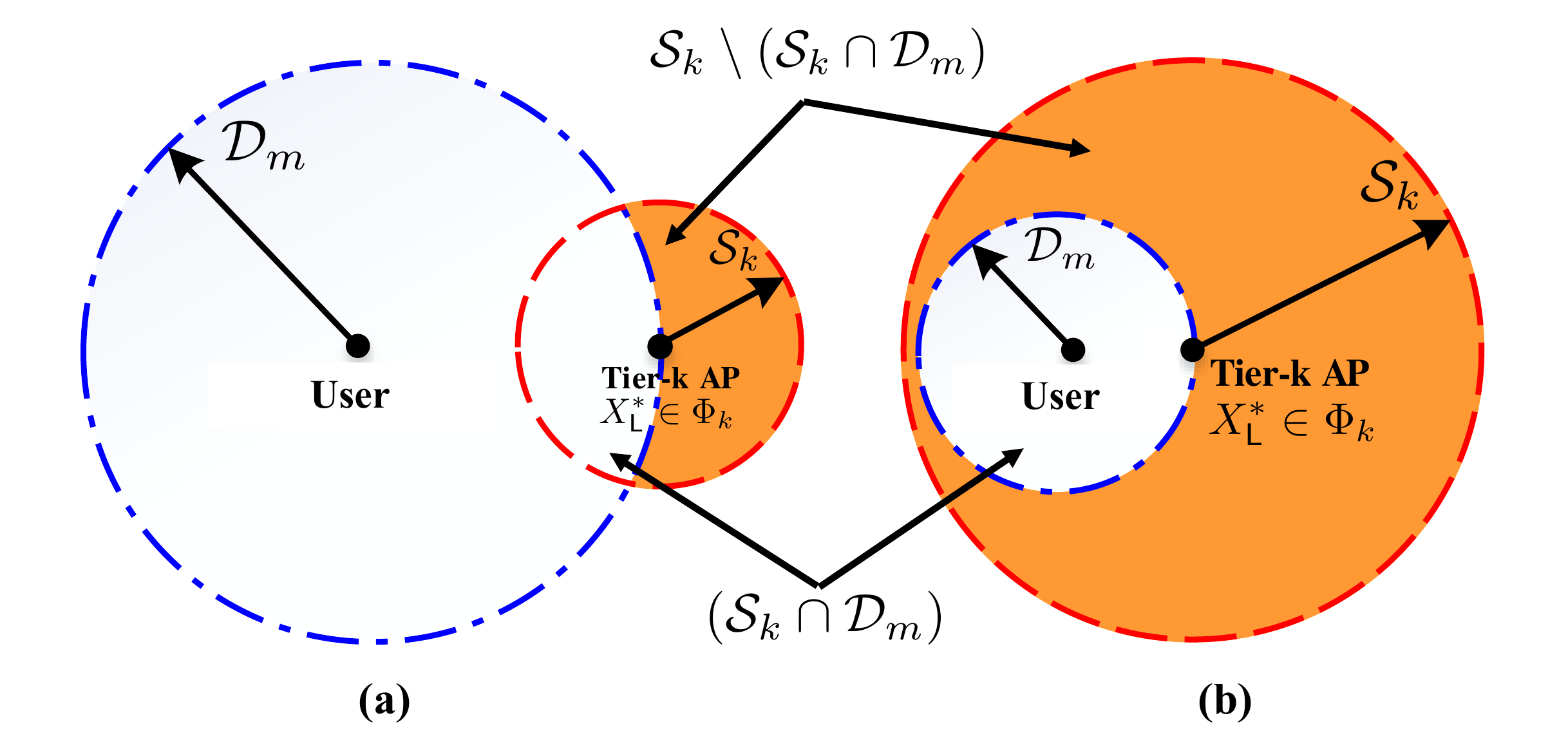}
	\caption{An illustration of regions $\mathcal{S}_k$, $\mathcal{D}_m$, $(\mathcal{S}_k\cap\mathcal{D}_m)$ and $\mathcal{S}_k\setminus(\mathcal{S}_k\cap\mathcal{D}_m)$ when a RAT-$\Ut$ user associates with its nearest AP: $\mathcal{S}_k$ is the sensing region of AP  $X^*_{\Lt}\in\Phi_k$  and $\mathcal{D}_m$ is the region in which there are no tier-$m$ APs. Sub-figure (a) shows the case that $\mathcal{D}_m$ is not enclosed by $\mathcal{S}_k$ and Sub-figure (b) shows the opposite case. Note that for the purpose of simple demonstration here $\mathcal{S}_k$ is shown as a circular region, however, it is not necessary to be circular in reality since the sensed signals usually suffer random channel impairments such as fading and shadowing.}
	\label{Fig:SensingRegionFig}
\end{figure}

To study the interactions while inconsistent random channel access protocols are operated in the HetNet, we consider the following channel access protocols for the two different RATs. For the APs in the first $K-1$ tiers, they are synchronized to simultaneously access the RAT-$\Lt$ channel if they have data to transmit to their users\footnote{To simplify the analyses in this work, we assume only one channel available in each of the RAT-$\Lt$ and RAT-$\Ut$ frequency bands. However, our analyses here can be extended to the multi-channel case; see our prior work in \cite{XDCHLLCWXZ16} for details.}. In other words, they share the RAT-$\Lt$ channel at the same time whenever they are transmitting. Before accessing the RAT-$\Ut$ channel, all the non-void APs have to contend it by using the slotted non-persistent opportunistic CSMA/CA protocol\footnote{To have a simpler interference model in the SIR model in Section \ref{Sec:NonCrossRAT}, the CSMA/CA protocol is assumed to be ``slotted".}. By adopting such a slotted opportunistic CSMA/CA protocol, all APs with channel gains in the RAT-$\Ut$ channel greater than some threshold are qualified and synchronized to access the RAT-$\Ut$ channel in the predesignated time slots. 

How a tier-$k$ AP can successfully access the RAT-$\Ut$ channel can be explained by using the illustration example shown in Fig. \ref{Fig:SensingRegionFig}. In Fig. \ref{Fig:SensingRegionFig}, $\mathcal{S}_k$ (red dash-circular disk) denotes the (random) sensing region of a tier-$k$ AP in which all transmitting activities of all other RAT-$\Ut$ APs can be detected by the tier-$k$ AP. Consider the noncrossing-RAT user association scenario and suppose RAT-$\Lt$ users always associate with their nearest AP. The user in the figure associates with its nearest AP $X^*_{\Lt}\in\Phi_k$ so that $\mathcal{D}_m$ (blue dash-circular disk) is the (random) region in which there are no tier-$m$ APs for all $m\in\mathcal{K}\setminus K$. In other words, the tier-$m$ APs that contend the RAT-$\Ut$ channel with $X^*_{\Lt}$  are only located in the shaded region $\mathcal{S}_k\setminus(\mathcal{S}_k\cap\mathcal{D}_m)$. Whereas the APs from the other $K-1$ tiers contending the channel can be located in the entire $\mathcal{S}_k$. Let $\delta(\mathcal{A})$ denote the Lebesgue measure (area) of set $\mathcal{A}$ and $A_{k,m}=\mathbb{E}[\delta(\mathcal{S}_k\setminus(\mathcal{S}_k\cap\mathcal{D}_m))]\mathds{1}(m\neq K)+\mathbb{E}[\delta(\mathcal{S}_k)]\mathds{1}(m=K)$ is the mean area within region $\mathcal{S}_k$ where the tier-$m$ APs are distributed. Using the definition of $A_{k,m}$, the probability of accessing the RAT-$\Ut$ channel for an RAT-$\Ut$ AP using opportunistic CSMA/CA can be found as shown in the following theorem if $F_Z(\cdot)$ and $f_Z(\cdot)$ denote the cumulative density function (CDF) and probability density function (pdf) of random variable $Z$, respectively.
\begin{theorem}\label{Thm:ChaAccProb}
First consider the scenario of noncrossing-RAT user association. Suppose a tier-$k$ AP has a sensing region of $\mathcal{S}_k$ and accesses the RAT-$\Ut$ channel by the opportunistic CSMA/CA protocol with channel gain threshold $\Delta>0$ and random backoff time $T_k\in[0,\tau_k]$ where $\tau_k\geq 0$ is the maximum backoff time of the tier-$k$ APs. Without loss of generality, assuming $0\leq\tau_K\leq\tau_{K-1}\leq\cdots\leq\tau_2\leq\tau_1$, the channel access probability of the tier-$k$ APs  is shown as
\begin{align}\label{Eqn:TransProb}
	\rho_k =\sum_{j=k}^{K}\int_{\tau_{j+1}}^{\tau_j} \exp\left(-\sum_{m=1}^{j}A_{k,m}p_m(1-\nu_{m})\tilde{\lambda}_m[F_{T_m}(\tau_j)-F_{T_m}(\tau_{j+1})]t\right) f_{T_j}(t) \dif t,
\end{align}
where $\tau_{K+1}\equiv 0$, $p_m\defn\mathbb{P}[H_mG_m^{-1}\geq\Delta]$ is the probability that the channel gain (fading gain $H_m\times$ shadowing gain $G^{-1}_m$) of an associated tier-$m$ AP is greater than $\Delta$ and $\tilde{\lambda}_m\defn\lambda_m\mathbb{E}\left[W^{-\frac{2}{\alpha}}_m\right]\mathbb{E}\left[W^{\frac{2}{\alpha}}_m\right]$. For the scenario of crossing-RAT user association, the channel access probability in \eqref{Eqn:TransProb} becomes
\begin{align}\label{Eqn:CrossRATTransProb}
\hat{\rho}_k =\sum_{j=k}^{K}\int_{\tau_{j+1}}^{\tau_j} \exp\left(-\sum_{m=1}^{j}\hat{A}_{k,m}p_m(1-\hat{\nu}_{m})\tilde{\lambda}_m[F_{T_m}(\tau_j)-F_{T_m}(\tau_{j+1})]t\right) f_{T_j}(t) \dif t,
\end{align} 
where $\hat{A}_{k,m}=\mathbb{E}[\delta(\mathcal{S}_k\setminus(\mathcal{S}_k\cap\mathcal{D}_m))]$. 
\end{theorem}
\begin{IEEEproof}
According to Reference \cite{FBBBL10}, the channel access probability of an AP using CSMA/CA can be shown as $e^{-tN}$ where $N$ is the average number of the contending APs and $t$ is the fixed backoff time for all APs. As a result, the probability that a tier-$k$ AP uses the opportunistic CSMA/CA with a random backoff time $T_k\in[0,\tau_k]$ to access the RAT-$\Ut$ channel can  be characterized by $\mathbb{E}\left[e^{-T_kN_k}\right]$ where $N_k$ is the mean number of the APs that are not void and have their channel gains higher than the threshold $\Delta$ in the sensing region $\mathcal{S}_k$ of a tier-$k$ AP as shown in Fig. \ref{Fig:SensingRegionFig}. If the noncrossing-RAT user association in \eqref{Eqn:NonCrossUserAssSch} is adopted, define  $\tilde{\Psi}_{\Lt}\defn \bigcup_{m=1}^{K-1} \tilde{\Phi}_m$ where $\tilde{\Phi}_m\defn\{\tilde{X}_{m,i}\in\mathbb{R}^2: \tilde{X}_{m,i}=W^{-\frac{1}{\alpha}}_{m,i}X_{m,i}, X_{m,i}\in\Phi_m,\forall i\in\mathbb{N}_+ \}$ is a homogeneous PPP of intensity $\lambda_m\mathbb{E}\left[W_m^{\frac{2}{\alpha}}\right]$ due to the results in Lemma \ref{Lem:StaDisUserAss} in Appendix \ref{App:LemmaNonCrossUserAss} so that the intensity of $\tilde{\Psi}_{\Lt}$ is $\sum_{m=1}^{K-1}\lambda_m\mathbb{E}\left[W^{\frac{2}{\alpha}}_m\right]$. Similarly, define $\tilde{\Psi}_{\Ut}\defn\{\tilde{X}_{K,i}\in\mathbb{R}^2: \tilde{X}_{K,i}=W^{-\frac{1}{\alpha}}_{K,i}X_{K,i}, X_{K,i}\in\Phi_K,\forall i\in\mathbb{N}_+ \}$ of intensity $\lambda_K\mathbb{E}\left[W^{\frac{2}{\alpha}}_K\right]$. Let $\tilde{X}^*_{\Lt}$ be the point in $\tilde{\Psi}_{\Lt}$ nearest to the origin so that $W^{-\frac{1}{\alpha}}_{\Lt}\|X^*_{\Lt}\|$ is almost surely equal to $\|\tilde{X}^*_{\Lt}\|$ since they have the same distribution. As such, the RAT-$\Lt$ typical user can be equivalently viewed to associate with AP $\tilde{X}^*_{\Lt}$ in probability and the sensing region of an AP in $\tilde{\Phi}_k$ becomes $\tilde{\mathcal{S}}_k$ whose mean area is $\mathbb{E}\left[\delta(\tilde{\mathcal{S}}_k)\right]=\mathbb{E}\left[\delta(\mathcal{S}_k)\right]\mathbb{E}\left[W^{-\frac{2}{\alpha}}_k\right]$.

According to the definition of $\mathcal{D}_m$ in Fig. \ref{Fig:SensingRegionFig}, similarly we define $\mathcal{\tilde{D}}_m$ as the region that does not contain any APs of $\tilde{\Phi}_m$ since $\tilde{X}^*_{\Lt}\in\tilde{\Psi}_{\Lt}$ is nearest to the origin. Accordingly, if $m\in\{1,\ldots,K-1\}$, then  $\tilde{\mathcal{S}}_k\setminus(\tilde{\mathcal{S}}_k\cap\tilde{\mathcal{D}}_m)$ represents the region in which the APs in $\tilde{\Phi}_m$ are distributed because no APs in $\tilde{\Phi}_m$ are distributed in $(\tilde{\mathcal{S}}_k\cap\tilde{\mathcal{D}}_m)$. That is, $A_{k,m}\defn\mathbb{E}\left[\delta(\mathcal{S}_k\setminus(\mathcal{S}_k\cap\mathcal{D}_m))\right]=\mathbb{E}\left[\delta(\tilde{\mathcal{S}}_k\setminus(\tilde{\mathcal{S}}_k\cap\tilde{\mathcal{D}}_m))\right]/\mathbb{E}\left[W^{-\frac{2}{\alpha}}_m\right]$. Hence, the average number of  the non-void APs of $\tilde{\Phi}_m$ in region $\tilde{\mathcal{S}}_k\setminus(\tilde{\mathcal{S}}_k\cap\tilde{\mathcal{D}}_m)$ is $(1-\nu_k)\lambda_k\mathbb{E}\left[W^{\frac{2}{\alpha}}_k\right]\mathbb{E}\left[\delta(\tilde{\mathcal{S}}_k\setminus(\tilde{\mathcal{S}}_k\cap\tilde{\mathcal{D}}_m))\right]=(1-\nu_k)\tilde{\lambda}_kA_{k,m}$ where $1-\nu_k=\mathbb{P}[V_{k,i}=1]$ is the non-void probability of the tier-$k$ APs. Whereas for $m=K$ non-void APs in the $K$th tier are distributed in the entire $\tilde{\mathcal{S}}_k$ so that $A_{k,K}=\mathbb{E}\left[\delta(\mathcal{S}_k)\right]=\mathbb{E}\left[\delta(\tilde{\mathcal{S}}_k)\right]/\mathbb{E}\left[W^{-\frac{2}{\alpha}}_K\right]$ and thus the average number of the tier-$m$ APs in $\tilde{\mathcal{S}}_k$ is $(1-\nu_m)\tilde{\lambda}_mA_{k,m}$. With the help of the illustration example shown in Fig. \ref{Fig:SensingRegionFig}, we can image that the APs from $\tilde{\Psi}_{\Lt}$ are only distributed in the ``shaded region" of $\tilde{\mathcal{S}}_k\setminus(\tilde{\mathcal{S}}_k\cap\tilde{\mathcal{D}}_m)$ and the APs from $\tilde{\Phi}_K$ can be distributed in the entire region of $\tilde{\mathcal{S}}_k$. For the scenario of crossing-RAT user association, $A_{k,m}$ is just equal to $\hat{A}_{k,m}=\mathbb{E}[\delta(\tilde{\mathcal{S}}_k\setminus(\tilde{\mathcal{S}}_k\cap\tilde{\mathcal{D}}_m))]/\mathbb{E}\left[W^{\frac{2}{\alpha}}_m\right]=\mathbb{E}[\delta(\mathcal{S}_k\setminus(\mathcal{S}_k\cap\mathcal{D}_m))]$ in that users can associate with any APs in the network. Note that if all $W_{k,i}$'s are equal to one (i.e., all users associate with their nearest AP) $\tilde{\mathcal{D}}_k$ reduces to $\mathcal{D}_k$ for all $k\in\mathcal{K}$, which is exactly the case shown in Fig. \ref{Fig:SensingRegionFig}. 

In addition, at each particular time point, the APs that are qualified to contend the RAT-$\Ut$ channel must have their random backoff time durations covering that time point, have channel gains greater than the threshold and are not void. Therefore, if considering the probability of random backoff time $T_m\in[\tau_{j+1},\tau_j]$, the intensity $(1-\nu_m)p_m\tilde{\lambda}_m$ is thinned by $[F_{T_m}(\tau_j)-F_{T_m}(\tau_{j+1})]$ so that the total average number of the APs contending the channel with a tier-$k$ AP during the time range of $[\tau_{j+1},\tau_j]$ is $\sum_{m=1}^j A_{k,m}p_m\tilde{\lambda}_m(1-\nu_m)[F_{T_m}(\tau_j)-F_{T_m}(\tau_{j+1})]$. Since $T_k\in[0,\tau_k]$ and $T_k=\bigcup^K_{j=k} [\tau_{j+1},\tau_j]$, we can show that $\mathbb{E}\left[e^{T_kN_k}\right]$is equal to the result in \eqref{Eqn:TransProb}. This completes the proof.
\end{IEEEproof}

Theorem \ref{Thm:ChaAccProb} essentially characterizes the inconsistence between the channel access protocols used by the APs in different tiers so that it is able to evaluate the channel access probabilities for the APs with different priorities. Also, it reveals that the  backoff time distribution significantly dominates the channel access probability and thus it plays a pivotal role on modeling the interference in the network. The following corollary shows the channel access probability of a tier-$k$ AP with a uniform-distributed random backoff time. 
\begin{corollary}
If the random backoff time $T_k$ of a tier-$k$ AP is uniformly distributed in $[0,\tau_k]$, then its channel access probability $\rho_k$ in \eqref{Eqn:TransProb} can be explicitly found as 
\begin{align}\label{Eqn:TransProbUniDis}
\rho_k=&\frac{1-e^{-\tau_K\sum\limits_{m=1}^{K}A_{k,m}\overline{\lambda}_{m,K}}}{\tau_k\sum_{m=1}^{K}A_{k,m}\overline{\lambda}_{m,K}}+\sum\limits_{j=k}^{K-1}\frac{e^{-\tau_{j}\sum_{m=1}^{j}A_{k,m}\overline{\lambda}_{m,j}}-e^{-\tau_{j+1}\sum_{m=1}^{j}A_{k,m}\overline{\lambda}_{m,j}}}{\tau_k\sum_{m=1}^jA_{k,m}\overline{\lambda}_{m,j}},
\end{align}
where $\overline{\lambda}_{m,j}\defn p_m(1-\nu_m)\tilde{\lambda}_m\left(\frac{\tau_j-\tau_{j+1}}{\tau_m}\right)$ for all  $m\in\mathcal{K}$. Furthermore, $\rho_k$ in \eqref{Eqn:TransProbUniDis} reduces to 
\begin{align}\label{Eqn:TransProbwithEqualTime}
\rho_k=\frac{1-\exp\left(-\tau\sum_{m=1}^{K}A_{k,m}p_m(1-\nu_m)\tilde{\lambda}_m\right)}{\tau\sum_{m=1}^{K}A_{k,m}p_m(1-\nu_m)\tilde{\lambda}_m}
\end{align}
if $\tau_1=\tau_2=\cdots=\tau_K=\tau$. The channel access probability in \eqref{Eqn:CrossRATTransProb} with uniform-distributed random backoff times can be easily obtained by replacing $A_{k,m}$ and $\nu_m$ in \eqref{Eqn:TransProbUniDis} with $\hat{A}_{k,m}$ and $\hat{\nu}_m$.
\end{corollary}
\begin{IEEEproof}
Since random backoff time $T_m$ are uniformly distributed over $[0,\tau_m]$, $F_{T_m}(\tau_j)-F_{T_m}(\tau_{j+1})=\frac{\tau_j-\tau_{j+1}}{\tau_m}$. First consider the last single term in \eqref{Eqn:TransProb} for $j=K$ and we have
$$\int_{0}^{\tau_K} \frac{1}{\tau_k}e^{-t\sum_{m=1}^{K}A_{k,m}\overline{\lambda}_{m,K}}\dif t=\frac{1-e^{-\tau_K\sum_{m=1}^{K}A_{k,m}\overline{\lambda}_{m,K}}}{\tau_k\sum_{m=1}^{K}A_{k,m}\overline{\lambda}_{m,K}},$$  
where $m$ is running from $1$ to $K$ because all random backoff times $T_m$'s cover the range of $[0,\tau_K]$. Now consider any other terms in \eqref{Eqn:TransProb} for $k\leq j\leq K-1$. Since $F_{T_m}(\tau)$ is known, it follows that
\begin{align*}
\int_{\tau_{j+1}}^{\tau_j} \frac{1}{\tau_k}e^{-t\sum_{m=1}^{j}A_{k,m}\overline{\lambda}_{m,j}}\dif t=\frac{e^{-\tau_j\sum_{m=1}^{j}A_{k,m}\overline{\lambda}_{m,j}}-e^{-\tau_{j+1}\sum_{m=1}^{j}A_{k,m}\overline{\lambda}_{m,j}}}{\tau_k\sum_{m=1}^jA_{k,m}\overline{\lambda}_{m,j}}
\end{align*}
and then summing up all terms in this equation from $j=k$ to $j=K$ results in \eqref{Eqn:TransProbUniDis}. If all $\tau_k$'s are the same and equal to $\tau$, all the terms in \eqref{Eqn:TransProbUniDis} vanish except the first term with $\tau_K=\tau$ and thereby \eqref{Eqn:TransProbwithEqualTime} is obtained.
\end{IEEEproof}

\begin{figure}[t!]
\centering
\includegraphics[width=4.5in,height=2.75in]{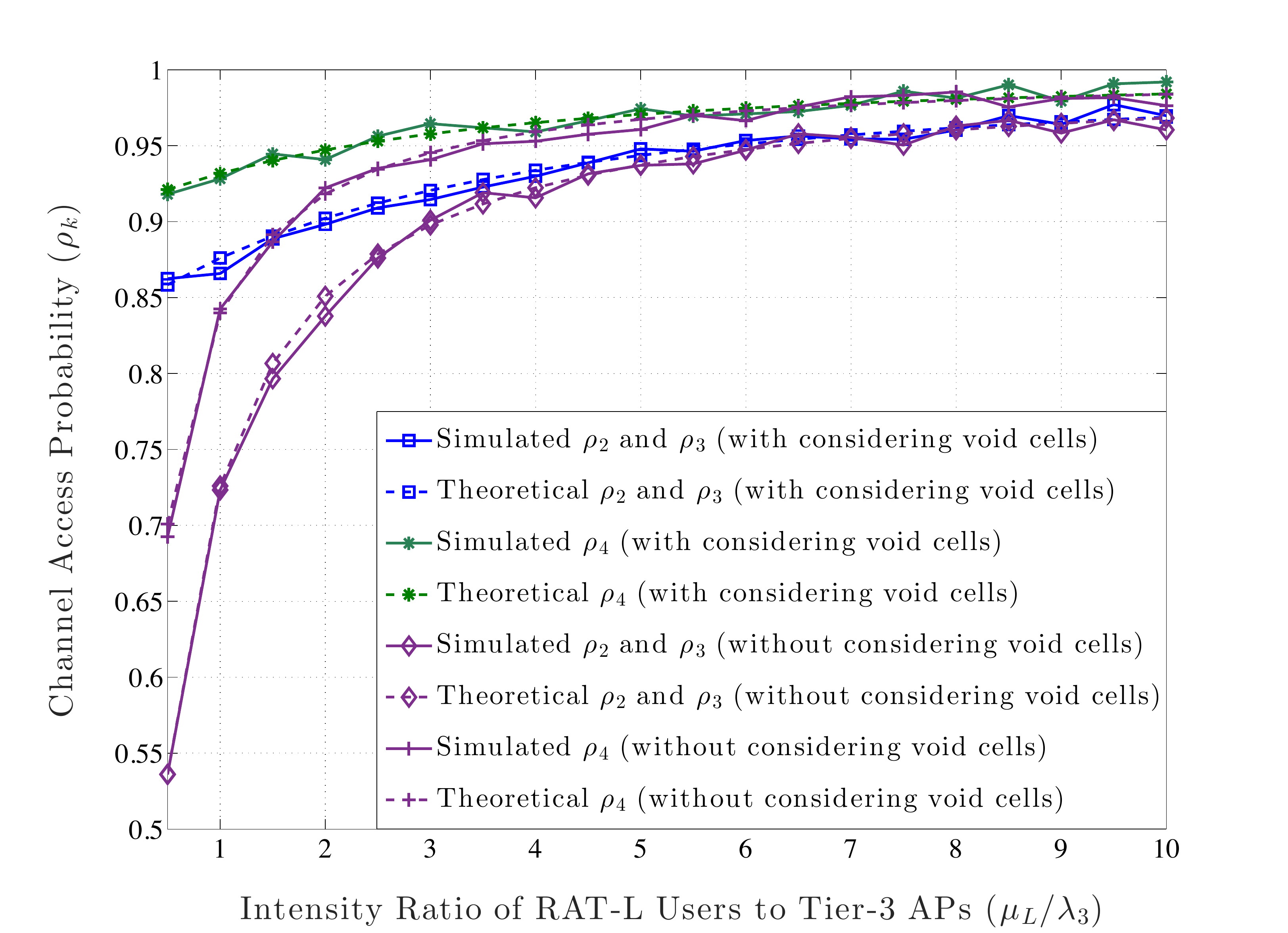}
\caption{Channel access probabilities of the RAT-$\Lt$ and RAT-$\Ut$ APs in a four-tier HetNet with the opportunistic CSMA/CA protocol and considering the void cell phenomenon. The network parameters for simulation are: $\lambda_2=1\times 10^{-5}$ APs/$m^2$, $\lambda_2:\lambda_3:\lambda_4=1:5:10$,  $\tau_1=\infty$, $\tau_2=\tau_3=2$, $\tau_4=1$, $\alpha=4$, $P_2=1$W, $P_3=0.5$W, $P_4=0.2$W, and $W_{k,i}=P_kG^{-1}_{k,i}$ where $G_k\sim\ln\mathcal{N}(0,3$ dB)  for all $k\in\{2,3,4\}$.}
\label{Fig:ChannelAccessProb}
\end{figure}

The simulation results of $\rho_k$ in \eqref{Eqn:TransProbUniDis}  are illustrated in Fig. \ref{Fig:ChannelAccessProb} for a four-tier HetNet by assuming the APs in the first tier are not allowed to contend the channel (i.e., $\tau_1=\infty$), $\mathcal{S}_2$, $\mathcal{S}_3$, $\mathcal{S}_4$ are a circular region of radius 30m. Other network parameters for simulation are given in the caption of Fig.  \ref{Fig:ChannelAccessProb}. As shown in the figure, the simulation results perfectly coincide with the theoretical results of $\rho_k$ in \eqref{Eqn:TransProbUniDis} and the APs in the second and third tiers have a lower channel access probability than those in the fourth tier due to their longer random backoff time ranges. Also, the APs in the case with considering the void cell phenomenon have a higher channel access probability than those without considering the void cell phenomenon especially when the intensities of APs are not much larger than the intensity of users. Hence, the void cell phenomenon should be also included in calculating the channel access probability especially in the scenario of dense AP deployment; however, this phenomenon is generally overlooked in prior works. Moreover, it is worth pointing out that opportunistic CSMA/CA definitely can improve the channel access probability of the APs with good channels if compared with traditional CSMA/CA even though Fig. \ref{Fig:ChannelAccessProb} does not illustrate this. In the following analysis, we will show how opportunistic CSMA/CA reduces interference to benefit the link coverage (probability) and capacity of the multi-RAT HetNet.

\section{Noncrossing-RAT User Association: Coexisting Coverage and Capacity}\label{Sec:NonCrossRAT}
In this section, we consider the scenario of noncrossing-RAT user association and would like to investigate the coverage and capacity problems in this scenario. To evaluate the coexisting transmission performance of the APs in the multi-RAT HetNet, we first need to define the SIR of RAT-$\Rt$ users by assuming the entire network is interference-limited. Without loss of generality, suppose a typical RAT-$\Lt$ user located in the origin and its serving AP can access the RAT-$\Ut$ channel. Therefore, its SIR, $\gamma_{\Lt}$ in the two RAT frequency bands, right after it associates with its serving AP, can be expressed as 
\begin{align}\label{Eqn:RATL-SIR}
\gamma_{\Lt}=\begin{cases}
\gamma_{\Lt_l}\defn\sum_{k=1}^{K-1}\frac{H_kG^{-1}_kP_k}{I_{\Lt_l}\|X^*_{\Lt}\|^{\alpha}}\mathds{1}(X^*_{\Lt}\in\Phi_k),&\text{for RAT-$\Lt$}\\
\gamma_{\Lt_u}\defn\sum_{k=1}^{K-1}\frac{H'_kG'^{-1}_kP_k}{I_{\Lt_u}\|X^*_{\Lt}\|^{\alpha}}\mathds{1}(X^*_{\Lt}\in\Phi_k), &\text{for RAT-$\Ut$}
\end{cases},
\end{align}
 respectively, where $I_{\Lt_l}\defn \sum_{k,i:X_{k,i}\in \bigcup_{k=1}^{K-1}\Phi_k\setminus X^*_{\Lt}}V_{k,i}P_kH_{k,i}G^{-1}_{k,i}\|X_{k,i}\|^{-\alpha}$ denotes the interference generated by the APs in the first $K-1$ tiers, $H_{k,i}$ and $G^{-1}_{k,i}$ denote the Rayleigh fading and shadowing gains of the RAT-$\Lt$ channel of AP $X_{k,i}$, respectively (all $H_{k,i}$'s are i.i.d. exponential random variables with unit mean and all $G^{-1}_{k,i}$'s are i.i.d. random variables for the same subscript $k$, and $H_{k,i}$'s as well as all $G^{-1}_{k,i}$'s are merely independent for different $k$'s), $I_{\Lt_u}\defn \sum_{k,i:X_{k,i}\in\Psi_{\Ut}\setminus X^*_{\Lt}}V_{k,i}P_kH'_{k,i}G'^{-1}_{k,i}\|X_{k,i}\|^{-\alpha}$ is the interference generated by set $\Psi_{\Ut}\subseteq\Phi$ consisting of all the APs accessing the RAT-$\Ut$ channel, $H'_{k,i}$ and $G'^{-1}_{k,i}$ are the Rayleigh fading and shadowing gains of the RAT-$\Ut$ channel of $X_{k,i}$, respectively (all $H'_{k,i}$'s are i.i.d. exponential random variables with unit mean and variance, all $G'^{-1}_{k,i}$'s are i.i.d. random variables for the same subscript $k$ and are independent for different $k$'s). Note that the channel fading and shadowing gains in $\gamma_{\Lt_l}$ and $\gamma_{\Lt_u}$ are also i.i.d. for the same tier (e.g., all $H_{k,i}$'s and $H'_{k,i}$'s are i.i.d. for all $k\in\mathcal{K}$ and $i\in\mathbb{N}_+$, $G^{-1}_{k,i}$ and $G'^{-1}_{k,i}$ are i.i.d. for the same subscript $k$.) and the interference in $I_{\Lt_u}$ contributed by the $K$th-tier APs is independent from $\|X^*_{\Lt}\|$ since $X^*_{\Lt}$ is from the first $K-1$ tiers. For a typical RAT-$\Ut$ user located at the origin, its SIR right after it associates with its serving AP can be expressed as
\begin{align}\label{Eqn:RAT-U_SIR}
\gamma_{\Ut}\defn \frac{H_KG^{-1}_KP_K}{I_{\Ut}\|X^*_{\Ut}\|^{\alpha}}
\end{align}
in which $I_{\Ut}\defn \sum_{m,i:X_{m,i}\in\Psi_{\Ut}\setminus X^*_{\Ut}}V_{m,i}P_{m}H_{m,i}G^{-1}_{m,i}\|X_{m,i}\|^{-\alpha}$ is the interference received by the RAT-$\Ut$ user and $G^{-1}_K\in\{G^{-1}_{K,i}, \forall i\in\mathbb{N}_+\}$ is the shadowing gain of $X^*_{\Ut}$. Also, note that the interference in $I_{\Ut}$ contributed by the APs in the first $K-1$ tiers is independent of $\|X^*_{\Ut}\|$ since $X^*_{\Ut}$ is from the $K$th tier. 

\subsection{Link Coverage, Coexisting Coverage and Their Limits}
By inheriting the concept of the coverage in single-RAT cellular networks, we define the \textit{coexisting coverage} for the multi-RAT heterogeneous network as follows.
\begin{definition}\label{Defn:CoeCovProb}
The link coverage of the RAT-$\Rt$ users in the multi-RAT HetNet is defined as
\begin{align}\label{Eqn:RATRCoverage}
\mathsf{P}_{\Rt}=\mathbb{P}[\gamma_{\Rt}\geq\theta],\quad \Rt\in\{\Lt,\Ut\},
\end{align}
where $\theta>0$ is the SIR threshold for successful decoding at the RAT-$\Rt$ users. According to the link coverage of RAT-$\Rt$ in \eqref{Eqn:RATRCoverage}, the coexisting coverage of the multi-RAT HetNet in the scenario of noncrossing-RAT user association is defined as
\begin{align}\label{Eqn:CoeCovProb}
\mathsf{P}_{\cov}\defn\sum_{k=1}^{K-1}\vartheta_k\mathbb{P}[\gamma_{\Lt_l}\geq\theta]+\vartheta_K\mathbb{P}[\gamma_{\Ut}\geq\theta]=\sum_{k=1}^{K-1}\vartheta_k\mathsf{P}_{\Lt_l}+\vartheta_K \mathsf{P}_{\Ut}.
\end{align} 
\end{definition}
\noindent The idea of proposing the coexisting coverage in Definition \ref{Defn:CoeCovProb} aims to provide an average coverage evaluation of a multi-RAT HetNet in which different RAT users are serviced at the same time.  Be aware that the coexisting coverage becomes the original coverage when the network just has a single RAT.  The following theorem gives the analytical results of the lower bounds on the coverage of each RAT and the coexisting coverage. 
\begin{theorem}\label{Thm:CoexistCoverage}
If the MMPA scheme is adopted, i.e., $W_{k,i}=P_kG^{-1}_{k,i}$ in \eqref{Eqn:NonCrossUserAssSch} for all $k\in\mathcal{K}$, the link coverage of the RAT-$\Lt$ users defined in \eqref{Eqn:RATRCoverage} is lower bounded by
	\begin{align}\label{Eqn:RatLCovProb}
	\mathsf{P}_{\Lt_l}\geq\frac{1}{1+\ell\left(\theta,\theta ;\frac{2}{\alpha}\right)\sum_{k=1}^{K-1}(1-\nu_k)\vartheta_k},
	\end{align}
	where $\vartheta_k=\lambda_kP^{\frac{2}{\alpha}}_k\mathbb{E}\left[G^{-\frac{2}{\alpha}}_k\right]/\sum_{m=1}^{K-1}\lambda_mP^{\frac{2}{\alpha}}_m\mathbb{E}\left[G^{-\frac{2}{\alpha}}_m\right]$ and function $\ell(\cdot,\cdot;\cdot)$ is defined as
	$$\ell(x,y;z)=x^z\left(\frac{\pi z}{\sin(\pi z)}-\int_{0}^{y^{-z}}\frac{\dif t}{1+t^{\frac{1}{z}}}\right).$$
	Whereas the link coverage of the RAT-$\Lt$ users in the RAT-$\Ut$ channel is $\mathsf{P}_{\Lt_u}\defn\mathbb{P}[\gamma_{\Lt_u}\geq\theta]$ and its lower bound is given by
	\begin{align}\label{Eqn:CovProbRATLU}
	\mathsf{P}_{\Lt_u}\geq \sum_{k=1}^{K-1} \mathbb{E}_{\frac{G'_k}{G_k}}\left\{\left(1+  \sum_{m=1}^{K}(1-\nu_m)\rho_mp_m\vartheta_m\mathbb{E}_{\frac{G'_m}{G_m}}\left[\ell\left(\frac{G'_mG'_k}{G_mG_k}\theta,\Theta_{k,m};\frac{2}{\alpha}\right)\right]\right)^{-1}\right\}\vartheta_k,
	\end{align}
	where $\Theta_{k,m}\defn  \frac{G'_mG'_k}{G_mG_k}\theta/\mathds{1}(m\neq K)$.
	For the link coverage of the RAT-$\Ut$ users, its lower bound is
	\begin{align}
	\mathsf{P}_{\Ut}\geq \left\{1+\vartheta^{-1}_K\sum_{k=1}^{K}\ell\left(\theta,\theta_k;\frac{2}{\alpha}\right)(1-\nu_k)\rho_kp_k\vartheta_k\right\}^{-1},\label{Eqn:RatUCovProb}
	\end{align}
where $\theta_k\defn \theta/\mathds{1}(k=K)$. Consequently, the lower bound on $\mathsf{P}_{\cov}$ can be acquired by plugging the lower bounds on $\mathsf{P}_{\Lt_l}$ and $\mathsf{P}_{\Ut}$ into \eqref{Eqn:CoeCovProb}. 
\end{theorem}
\begin{IEEEproof}
See Appendix \ref{App:ProofCoexistCoverage}.
\end{IEEEproof}

The lower bound on $\mathsf{P}_{\Lt}$ in \eqref{Eqn:RatLCovProb} that includes the void probability impact represents $\mathsf{P}_{\Lt}$ in the worse case, which means $\mathsf{P}_{\Lt}$ cannot lower below this limit. As a result of the location correlations between non-void APs induced by user association, the closed-form  $\mathsf{P}_{\Lt}$ cannot be found and $\mathsf{P}_{\Lt}$ will approach to its lower bound as the intensity of the RAT-$\Lt$ users goes to infinity (i.e., $\nu_k$ goes to zero.). That is, we have the following lowest limit on $\mathsf{P}_{\Lt_l}$:
\begin{align}
\underline{\mathsf{P}_{\Lt_l}}\defn\lim_{\mu_{\Lt}\rightarrow\infty} 	\mathsf{P}_{\Lt_l}=\frac{1}{1+\ell\left(\theta,\theta;\frac{2}{\alpha}\right)},
\end{align}
which does not depend on the AP intensity and it coincides with the link coverage of users in a single-RAT Poisson cellular network that overlooks the void cell phenomenon\cite{JGAFBRKG11}. Whereas $\mathsf{P}_{\Lt_l}\rightarrow 1$ as $\mu_{\Lt}/\lambda_k\rightarrow 0$, which is obvious because the intensity of APs is extremely large relative to the user intensity so that void probabilities are approaching to one. Hence, here we can make an important conclusion that \textit{the link coverage of the RAT-$\Lt$ users (or single-RAT link coverage) indeed depends on the intensities of users and APs and it increases as the user intensity goes to zero and/or AP intensity goes to infinity}. Hence, deploying more APs improves the link coverage. Although $\mathsf{P}_{\Lt_l}$ cannot be found in closed-form, its lower bound in \eqref{Eqn:RatLCovProb} is in general very tight since the correlations between the non-void APs are fairly weak as long as the user intensity is not extremely smaller than the total equivalent intensity of RAT-$\Lt$ APs, $\sum_{k=1}^{K-1} \lambda_kP^{\frac{2}{\alpha}}_k\mathbb{E}\left[G^{-\frac{2}{\alpha}}_k\right]$. In other words, the lower bound in \eqref{Eqn:RatLCovProb} usually can provide a good estimate of the link coverage in a typical network and this will be numerically verified in Section \ref{Sec:Simulation}. Also, if all $P_kG^{-1}_k$'s are i.i.d. and this leads to $\nu_1=\cdots=\nu_{K}=\nu$, we have
\begin{align}\label{Eqn:RATLCovProbLowLimit}
\mathsf{P}_{\Lt_l}\approx \frac{1}{1+(1-\nu)\ell\left(\theta,\theta;\frac{2}{\alpha}\right)},
\end{align}
which corresponds to the coverage in the case of nearest AP association and this result shows that the single-RAT link coverages found in prior works, such as \cite{JGAFBRKG11,HSJYJSXPJGA12}, are not precisely correct owing to overlooking the fact of cell voidness. According to the lower bound on the link coverage of the RAT-$\Lt$ users in the RAT-$\Ut$ channel in \eqref{Eqn:CovProbRATLU},  we can learn that large shadowing power significantly reduces $P_{\Lt_u}$ since the RAT-$\Ut$ channel variations due to shadowing are not explored by the RAT-$\Lt$ users and opportunistic CSMA/CA significantly increases $P_{\Lt_u}$ because some fraction of all APs are refrained from accessing the RAT-$\Ut$ channel and the interference in the RAT-$\Ut$ channel could be less than that in the RAT-$\Lt$ channel. The lowest limit on $P_{\Lt_u}$ is given by
\begin{align}
\underline{P_{\Lt_u}}\defn \lim_{\mu_{\Lt},\tau^{-1}_k\rightarrow\infty} P_{\Lt_u}=\sum_{k=1}^{K-1} \mathbb{E}_{\frac{G'_k}{G_k}}\left\{\left(1+  \sum_{m=1}^{K}p_m\vartheta_m\mathbb{E}_{\frac{G'_m}{G_m}}\left[\ell\left(\frac{G'_mG'_k}{G_mG_k}\theta,\Theta_{k,m};\frac{2}{\alpha}\right)\right]\right)^{-1}\right\}\vartheta_k.\label{Eqn:LowLimitRatLuCovProb}
\end{align}
 
The lower bound on $\mathsf{P}_{\Ut}$ is the achievable lowest limit since it is found by assuming the joint thinning point process of all APs still form a PPP after they perform user association and opportunistic CSMA/CA and such an assumption is the worse-case point process of inducing the largest interference. Accordingly, as the intensity of all users goes to infinity the lower bound on $\mathsf{P}_{\Ut}$ decreases and becomes
\begin{align}
\lim_{\mu_{\Ut},\mu_{\Lt}\rightarrow\infty}\mathsf{P}_{\Ut}\geq \left\{1+\frac{1}{\vartheta_K} \left[\ell\left(\theta,\theta;\frac{2}{\alpha}\right)\rho_Kp_K+\frac{2\pi\theta^{\frac{2}{\alpha}}\sum_{k=1}^{K-1}\rho_kp_k\vartheta_k}{\alpha\sin(2\pi/\alpha)}\right]\right\}^{-1}.\label{Eqn:LowLimitRatUCovProb1}
\end{align}
In this case, $\mathsf{P}_{\Ut}$ does not converge to its lower limit, like the case in $\mathsf{P}_{\Lt_l}$, is because all APs accessing the RAT-$\Ut$ channel form an Mat\'{e}rn hard-core point process (MHPP) due to CSMA/CA. As a result, if all APs can arbitrarily access the RAT-$\Ut$ channel as long as their channel gains are higher than the predesignated threshold $\Delta$, $\mathsf{P}_{\Ut}$ will approach to a limit given by
\begin{align}
\underline{\mathsf{P}_{\Ut}}\defn\lim_{\mu_{\Ut},\mu_{\Lt},\tau^{-1}_k\rightarrow\infty}\mathsf{P}_{\Ut}=\left\{1+\frac{1}{\vartheta_K}\left[\ell\left(\theta,\theta;\frac{2}{\alpha}\right)p_K+\frac{2\pi\theta^{\frac{2}{\alpha}}\sum_{k=1}^{K-1}p_k\vartheta_k}{\alpha\sin(2\pi/\alpha)}\right]\right\}^{-1}, \label{Eqn:LowLimitRatUCovProb2}
\end{align}
which is \textit{the lowest limit that can be achieved by} $\mathsf{P}_{\Ut}$. Moreover, we also have
\begin{align}
\mathsf{P}_{\Ut}\geq \left\{1+\frac{(1-\nu)}{\vartheta_K}\left[\ell\left(\theta,\theta;\frac{2}{\alpha}\right)\rho_Kp_K+\frac{2\pi\sum_{k=1}^{K-1}\rho_kp_k\vartheta_k}{\alpha\sin(2\pi/\alpha)}\right]\right\}^{-1}\label{Eqn:LowLimitRatUCovProb3}
\end{align}
if all $\nu_k$'s are the same and equal to $\nu$. By comparing \eqref{Eqn:RATLCovProbLowLimit} with \eqref{Eqn:LowLimitRatUCovProb3}, we can see that the lower bound on $P_{\Ut}$ is lower than that on $\mathsf{P}_{\Lt_l}$ since the RAT-$\Lt$ APs also have some opportunities to access the RAT-$\Ut$ channel and contribute some interference in the channel. Hence, the interplay between two link coverages $\mathsf{P}_{\Lt_l}$ and $\mathsf{P}_{\Ut}$ exists as long as $\mu_{\Lt}$ does not go to infinity. For example, increasing the intensities of the RAT-$\Lt$ APs makes $\mathsf{P}_{\Lt_l}$ and $\mathsf{P}_{\Ut}$ both reduce since more interference comes from the RAT-$\Lt$ APs. Hence, coexisting coverage $\mathsf{P}_{\cov}$ in \eqref{Eqn:CoeCovProb} that appropriately combines $\mathsf{P}_{\Lt_l}$ and $\mathsf{P}_{\Ut}$ can characterize their interplay as well as provide an overall coverage evaluation in a multi-RAT HetNet. The lowest limit on $\mathsf{P}_{\cov}$  can be obtained in closed-form as
\begin{align}
\underline{\mathsf{P}_{\cov}}=\sum_{k=1}^{K-1}\vartheta_k\underline{\mathsf{P}_{\Lt_l}}+\vartheta_K\underline{\mathsf{P}_{\Ut}}
\end{align}
as $\mathsf{P}_{\Lt_l}$ and $\mathsf{P}_{\Ut}$ converge to their lowest limits.

\subsection{Mean Spectrum Efficiency, Coexisting Network Capacity and Their Limits}
The mean spectrum efficiency (ergodic capacity per unit bandwidth) of the RAT-$\Lt$ users can be written as
\begin{align}\label{Eqn:DefnMeanCapRATL}
\mathsf{C}_{\Lt} = \mathbb{E}\left[\log_2\left(1+\gamma_{\Lt_l}\right)+\left(\sum_{k=1}^{K-1}\rho_kp_k\vartheta_k\right)\log_2\left(1+\gamma_{\Lt_u}\right)\right],\,\,\text{(bps/Hz)}
\end{align}
where the term $\sum_{k=1}^{K-1}\rho_kp_k\vartheta_k$ can be interpreted as the fraction of the total time that the RAT-$\Lt$ APs can access the RAT-$\Ut$ channel in the long term sense. Similarly, the mean spectrum efficiency of the RAT-$\Ut$ users can be expressed as follows
\begin{align}\label{Eqn:DefnMeanCapRATU}
\mathsf{C}_{\Ut}=\rho_Kp_K\mathbb{E}\left[\log_2\left(1+\gamma_{\Ut}\right)\right],\,\,\text{(bps/Hz)}.
\end{align}
To evaluate how much traffic can be carried by the multi-RAT HetNet, the \textit{coexisting network capacity} of the multi-RAT HetNet is proposed and defined as follows.
\begin{definition}\label{Defn:CoeNetCapa}
The coexisting network capacity of the multi-RAT HetNet, denoted by $\mathsf{C}_{\cov}$, is defined as the total sum of the mean successful spectrum efficiencies of all different RATs per unit area. Specifically, it can be expressed in terms of $\mathsf{C}_{\Lt}$ and $\mathsf{C}_{\Ut}$  as
\begin{align}
\mathsf{C}_{\cov}\defn \sum_{k=1}^{K-1}\lambda_k(1-\nu_k) \mathsf{P}_{\Lt_l}\mathsf{C}_{\Lt}+\lambda_K(1-\nu_K)\mathsf{P}_{\Ut}\mathsf{C}_{\Ut},\quad (\text{bps/Hz/m}^2). \label{Eqn:CovNetCap}
\end{align}
\end{definition}
\noindent The coexisting network capacity is essentially the metric of the mean successful area spectrum efficiency of the multi-RAT HetNet and especially it characterizes the void cell phenomenon that is hardly studied in the prior works on wireless network capacity. Note that $\mathsf{C}_{\cov}$ depends on what kind of user association is adopted since user association schemes affect the void cell probability, link coverage and mean spectrum efficiency. 

Since the exact results of $\mathsf{C}_{\Lt}$ and $\mathsf{C}_{\Ut}$ cannot be found due to the fact that the transmitting RAT-$\Lt$ and RAT-$\Ut$ APs are no longer PPPs any more, we resort to deriving their ``maximum'' lower bounds based on the link coverage results in Theorem \ref{Thm:CoexistCoverage} as shown in the following corollary.
\begin{corollary}\label{Thm:LowerBoundMeanRate}
If the MMPA scheme is adopted, the lower bound on $\mathsf{C}_{\Lt}$ is given by
\begin{align}\label{Eqn:RATLMeanRate}
\mathsf{C}_{\Lt}\geq  &\int_{0}^{\infty}\frac{\dif\theta}{(\ln 2)(1+\theta)\left[1+\ell\left(\theta,\theta;\frac{2}{\alpha}\right)\sum_{k=1}^{K-1}(1-\nu_k)\vartheta_k\right]}+\sum_{k=1}^{K-1}\int_{0}^{\infty}\frac{\left(\sum_{m=1}^{K-1}\rho_mp_m\vartheta_m\right)\vartheta_k}{(\ln 2)(1+\theta)}\nonumber\\
& \mathbb{E}_{\frac{G'_k}{G_k}}\left\{\left(1+\sum_{m=1}^{K}(1-\nu_m)\rho_mp_m\vartheta_m\mathbb{E}_{\frac{G'_m}{G_m}}\left[\ell\left(\frac{G'_kG'_m}{G_kG_m}\theta,\Theta_{k,m};\frac{2}{\alpha}\right)\right]\right)^{-1}\right\}\dif\theta
\end{align}
and the lower bound on $\mathsf{C}_{\Lt}$ is shown as
\begin{align}\label{Eqn:RATUMeanRate}
\mathsf{C}_{\Ut}\geq \frac{1}{\ln 2}\int_{0}^{\infty}\frac{\rho_KP_K\dif\theta}{(1+\theta)\left(1+ \vartheta^{-1}_K\sum_{k=1}^{K}\ell\left(\theta,\theta_k;\frac{2}{\alpha}\right)\rho_kp_k(1-\nu_k)\vartheta_k\right)}.
\end{align}
The lower bound on coexisting network capacity $\mathsf{C}_{\cov}$ can be found by substituting the lower bounds in \eqref{Eqn:RATLMeanRate} and \eqref{Eqn:RATUMeanRate} into \eqref{Eqn:CovNetCap}. 
\end{corollary}
\begin{IEEEproof}
According to $\mathsf{C}_{\Lt}$ in \eqref{Eqn:DefnMeanCapRATL}, $\mathsf{C}_{\Lt}$ can be equivalently expressed as
\begin{align*}
\mathsf{C}_{\Lt} &=\frac{1}{\ln 2}\left\{\int_{0}^{\infty}\mathbb{P}\left[\gamma_{\Lt_l}\geq e^x-1\right]\dif x+\left(\sum_{k=1}^{K-1}\rho_kp_k\vartheta_k\right)\int_{0}^{\infty}\mathbb{P}\left[\gamma_{\Lt_u}\geq e^x-1\right]\dif x\right\}\\
&=\frac{1}{\ln 2}\left\{\int_{0}^{\infty}\frac{\mathsf{P}_{\Lt_l}(\theta)}{1+\theta}\dif\theta+\left(\sum_{k=1}^{K-1}\rho_kp_k\vartheta_k\right)\int_{0}^{\infty}\frac{\mathsf{P}_{\Lt_u}(\theta)}{1+\theta}\dif \theta\right\}
\end{align*}
and then substituting the lower bounds in \eqref{Eqn:RatLCovProb} and \eqref{Eqn:CovProbRATLU} into the result of $\mathsf{C}_{\Lt}$ in above yields the lower bound in \eqref{Eqn:RATLMeanRate}. The lower bound on $\mathsf{C}_{\Ut}$ in \eqref{Eqn:RATUMeanRate} can be derived by following the same steps of deriving the lower bound on $\mathsf{C}_{\Lt}$. 
\end{IEEEproof}
Although only the lower bounds on $\mathsf{C}_{\Lt}$ and $\mathsf{C}_{\Ut}$ can be obtained in Theorem \ref{Thm:LowerBoundMeanRate}, they are actually very tight in some cases. For example, if all APs does not use CSAM/CA and can access the RAT-$\Ut$ channel once their channel gains are higher than the threshold, $\mathsf{C}_{\Lt}$ and $\mathsf{C}_{\Ut}$ can be accurately approximated by their lower bounds, i.e.,
\begin{align}
\lim_{\tau_k\rightarrow 0}\mathsf{C}_{\Lt}\approx &\int_{0}^{\infty}\frac{1/(\ln 2)(1+\theta)}{\left[1+\ell\left(\theta,\theta;\frac{2}{\alpha}\right)\sum_{k=1}^{K-1}(1-\nu_k)\vartheta_k\right]}\dif\theta\nonumber\\
&+\sum_{k=1}^{K-1}\int_{0}^{\infty}\mathbb{E}_{\frac{G'_k}{G_k}}\left[ \frac{\left(\sum_{m=1}^{K-1}p_m\vartheta_m\right)\vartheta_k/(\ln 2)(1+\theta)}{1+   \sum_{m=1}^{K}(1-\nu_m)p_m\vartheta_m\mathbb{E}_{\frac{G'm}{G_m}}\left[\ell\left(\frac{G'_kG'_m}{G_kG_m}\theta,\Theta_{k,m};\frac{2}{\alpha}\right)\right]}\right]\dif\theta\label{Eqn:AppRATLMeanRate}
\end{align}
and
\begin{align}
\lim_{\tau_k\rightarrow 0}\mathsf{C}_{\Ut}\approx\frac{1}{\ln 2}\int_{0}^{\infty}\frac{\dif\theta}{(1+\theta)\left(1+\sum_{k=1}^{K}\ell\left(\theta,\theta_k;\frac{2}{\alpha}\right)p_k(1-\nu_k)\vartheta_k/\vartheta_K\right)}. \label{Eqn:AppRATUMeanRate}
\end{align}
$\mathsf{C}_{\Lt}$ and $\mathsf{C}_{\Ut}$ do not exactly converge to their lower bounds in \eqref{Eqn:AppRATLMeanRate} and \eqref{Eqn:AppRATUMeanRate} because the non-void APs are no longer PPPs even though their location correlations are fairly weak in general. Accordingly, when user intensities go to infinity $\mathsf{C}_{\Lt}$ and $\mathsf{C}_{\Ut}$ will exactly reduce to their lowest limits respectively, $\underline{\mathsf{C}_{\Lt}}\defn\lim_{\mu_{\Lt},\mu_{\Ut},\tau^{-1}_k\rightarrow\infty}\mathsf{C}_{\Lt}$ and $\underline{\mathsf{C}_{\Ut}}\defn\lim_{\mu_{\Lt},\mu_{\Ut},\tau^{-1}_k\rightarrow \infty}\mathsf{C}_{\Ut}$ given by
\begin{align}
\underline{\mathsf{C}_{\Lt}}=&\sum_{k=1}^{K-1}\int_{0}^{\infty} \mathbb{E}_{\frac{G'_k}{G_k}}\left[\frac{\left(\sum_{m=1}^{K-1}p_m\vartheta_m\right)\vartheta_k/(\ln 2)(1+\theta)}{1+  \sum_{m=1}^{K}p_m\vartheta_m\mathbb{E}_{\frac{G'_m}{G_m}}\left[\ell\left(\frac{G'_kG'_m}{G_kG_m}\theta,\Theta_{k,m};\frac{2}{\alpha}\right)\right]}\right]\dif\theta +\int_{0}^{\infty}\frac{1/(\ln 2)(1+\theta)}{\left[1+ \ell\left(\theta,\theta;\frac{2}{\alpha}\right)\right]}\dif\theta,\\
\underline{\mathsf{C}_{\Ut}}= &\frac{1}{\ln 2}\int_{0}^{\infty}\frac{\dif\theta}{(1+\theta)\left(1+\sum_{k=1}^{K}\ell\left(\theta,\theta_k;\frac{2}{\alpha}\right)p_k\vartheta_k/\vartheta_K\right)}.
\end{align}
Under the same situation, the coexisting network capacity reduces to its lowest limit given by
\begin{align}
\underline{C_{\cov}}\defn\lim_{\mu_{\Lt},\mu_{\Ut},\tau^{-1}_k\rightarrow\infty}\mathsf{C}_{\cov}=\sum_{k=1}^{K-1}\lambda_k\underline{\mathsf{P}_{\Lt_l}}\,\underline{\mathsf{C}_{\Lt}}+\lambda_K\underline{\mathsf{P}_{\Ut}}\,\underline{\mathsf{C}_{\Ut}}. 
\end{align}
The closed-form $\underline{C_{\cov}}$ can be obtained based on the previous results of $\underline{\mathsf{P}_{\Lt_l}}$, $\underline{\mathsf{C}_{\Lt}}$, $\underline{\mathsf{P}_{\Ut}}$ and $\underline{\mathsf{C}_{\Ut}}$, which is an important result that not only shows the lowest limit on the network capacity, but also indicates how to increase the network capacity by deploying the APs in each tier with a proper intensity.  

\section{Crossing-RAT User Association: Coexisting Coverage and Capacity}\label{Sec:CrossRAT}
In this section, we consider the other scenario that all users can associate any AP in the $K$ tiers no matter which RAT is adopted by the APs. Namely, users associate with an AP by using the user association scheme in \eqref{Eqn:CrossUserAss}. Consider a typical user located in the origin and it associates with an AP in the first $K-1$ tiers. Its SIR in the RAT-$\Lt$ frequency band on the distance of the associated AP in \eqref{Eqn:CrossUserAssDis}, denoted by $\hat{\gamma}_{\Lt}$, like the noncrossing-RAT case, is similarly expressed as
\begin{align}\label{Eqn:CrossRAT-L_SIR}
\hat{\gamma}_{\Lt}=\begin{cases}
\hat{\gamma}_{\Lt_l}=\sum_{k=1}^{K-1}\frac{P_kH_kG_k^{-1}}{\hat{I}_{\Lt_l}\|X^*\|^{\alpha}}\mathds{1}(X^*\in\Phi_k),&\text{for RAT-$\Lt$}\\
\hat{\gamma}_{\Lt_u}=\sum_{k=1}^{K-1}\frac{P_kH'_kG_k'^{-1}}{\hat{I}_{\Lt_u}\|X^*\|^{\alpha}}\mathds{1}(X^*\in\Phi_k),&\text{for RAT-$\Ut$}
\end{cases},
\end{align}
where $\hat{I}_{\Lt_l}=\sum_{X_{m,i}\in\bigcup_{m=1}^{K-1} \Phi_m\setminus X^*}V_{m,i}P_mH_{m,i}G^{-1}_{m,i}\|X_{m,i}\|^{-\alpha}$ is the interference in the RAT-$\Lt$ channel and $\hat{I}_{\Lt_u}=\sum_{X_{m,i}\in\Psi_{\Ut}\setminus X^*}V_{m,i}P_mH'_{m,i}G'^{-1}_{m,i}\|X_{m,i}\|^{-\alpha}$ is the interference in the RAT-$\Ut$ channel. Similarly, if the typical user associates with a tier-$K$AP, its SIR in the RAT-$\Ut$ frequency band can be written as
\begin{align}\label{Eqn:CrossRAT-U_SIR}
\hat{\gamma}_{\Ut}\defn \frac{H_KG^{-1}_KP_K}{\hat{I}_{\Ut}\|X^*\|^{\alpha}},
\end{align}
where $\hat{I}_{\Ut}=\sum_{X_{m,i}\in\Psi_{\Ut}\setminus X^*}V_{m,i}P_mH_{m,i}G^{-1}_{m,i}\|X_{m,i}\|^{-\alpha}$ and $X^*\in\Phi_K$. The coverage of the typical user in the RAT-$\Rt$ frequency band is also defined as $\hat{\mathsf{P}}_{\Rt}\defn \mathbb{P}[\hat{\gamma}_{\Rt}\geq \theta]$ for $\Rt\in\{\Lt,\Ut\}$.
According to the definitions of the crossing-RAT SIRs, we can use them to derive the coexisting coverage, mean spectrum efficiency and coexisting network capacity as shown in the following subsections.

\subsection{Link Coverage, Coexisting Coverage and Their Limits}
According to the coexisting coverage defined in \eqref{Eqn:CoeCovProb} for the crossing-RAT scenario, we also define the coexisting coverage in the crossing-RAT scenario based on the link coverages in the two RAT channels as 
\begin{align}\label{Eqn:CrossRATCoeCovProb}
\hat{\mathsf{P}}_{\cov}=\sum_{k=1}^{K-1}\vartheta_k\hat{\mathsf{P}}_{\Lt}+\vartheta_K\hat{\mathsf{P}}_{\Ut}
\end{align} 
because the probability that a user associates with a tier-$k$ AP is $\vartheta_k$ and the total probability that the user associates with an RAT-$\Lt$ AP is $\sum_{k=1}^{K-1}\vartheta_k$. The following theorem gives the lower bounds on the link coverages in the two frequency and the coexisting coverage for the MMPA scheme. 
\begin{theorem}\label{Thm:CrossRATCoeCoverage}
Suppose all users adopt the MMPA scheme to associate their APs from all $K$ tiers. If a user associates with an AP in the first $K-1$ tiers, the lower bounds on its link coverages in the RAT-$\Lt$ and RAT-$\Ut$ channels can be shown as
\begin{align}
\hat{\mathsf{P}}_{\Lt_l} &\geq   \frac{1}{1+\ell\left(\theta,\theta;\frac{2}{\alpha}\right)\sum_{m=1}^{K-1}(1-\hat{\nu}_m)\hat{\vartheta}_m},\label{Eqn:NonCrossRATCovProbRATLl}\\
\hat{\mathsf{P}}_{\Lt_u} &\geq \sum_{k=1}^{K-1} \mathbb{E}_{\frac{G'_k}{G_k}}\left\{\left(1+ \sum_{m=1}^{K}(1-\hat{\nu}_m)\hat{\rho}_mp_m\hat{\vartheta}_m\mathbb{E}_{\frac{G'_m}{G_m}}\left[\ell\left(\hat{\Theta}_{k,m},\hat{\Theta}_{k,m};\frac{2}{\alpha}\right)\right]\right)^{-1}\right\}\vartheta_k, \label{Eqn:CrossRATCovProbRATLu}
\end{align}
where $\hat{\vartheta}_m=\lambda_mP^{\frac{2}{\alpha}}_m\mathbb{E}\left[G^{-\frac{2}{\alpha}}_m\right]/\sum_{k=1}^{K}\lambda_kP^{\frac{2}{\alpha}}_k\mathbb{E}\left[G^{-\frac{2}{\alpha}}_k\right]$, $\hat{\nu}_m$ is given in \eqref{Eqn:VoidProbCrossRAT}, $\hat{\rho}_m$ is given in \eqref{Eqn:CrossRATTransProb}, and $\hat{\Theta}_{k,m}=\frac{G'_kG'_m}{G_kG_m}\theta\mathds{1}(m\neq K)+\theta\mathds{1}(m=K)$, for $k\in\{1,2,\ldots,K-1\}$. On the other hand, if the user associates with an AP in the $K$th tier, the user's link coverage is lower bounded by
\begin{align}\label{Eqn:CrossRATCovRATU}
\hat{\mathsf{P}}_{\Ut}\geq \left\{1+\sum_{m=1}^{K}(1-\hat{\nu}_m)\hat{\rho}_mp_m\hat{\vartheta}_m\mathbb{E}\left[\ell\left(\hat{\Theta}_{K,m},\hat{\Theta}_{K,m};\frac{2}{\alpha}\right)\right]\right\}^{-1},
\end{align}
where $\hat{\Theta}_{K,m}=\frac{\theta}{G_m}\mathds{1}(m\neq K)+\theta\mathds{1}(m=K)$. The lower bound on $\hat{\mathsf{P}}_{\cov}$ can be obtained by substituting \eqref{Eqn:CrossRATCovProbRATLu} and \eqref{Eqn:CrossRATCovRATU} into \eqref{Eqn:CrossRATCoeCovProb}.
\end{theorem}
\begin{IEEEproof}
The proof is similar to the proof of Theorem \ref{Thm:CoexistCoverage} and omitted due to limited space.
\end{IEEEproof}

In general, the coverages in Theorem \ref{Thm:CrossRATCoeCoverage} are greater than those in Theorem \ref{Thm:CoexistCoverage} in that users have one more teir of APs to select so that their SIR increases since they have a better opportunity to associate with an AP with a higher channel gain and the void probabilities increases as well under the same user intensity.  For example, the lower bound on $P_{\Lt}$ is larger than the lower bound on $\hat{P}_{\Lt}$ since $\nu_k$ is smaller than $\hat{\nu}_k$ whereas $\vartheta_k$ is greater than $\hat{\vartheta}_k$.  The lowest limits on $\hat{\mathsf{P}}_{\Lt_l}$, $\hat{\mathsf{P}}_{\Lt_u}$ and $\hat{\mathsf{P}}_{\Ut}$ are
\begin{align}
\underline{\hat{\mathsf{P}}_{\Lt_l}}&=\lim_{\mu\rightarrow\infty}\hat{\mathsf{P}}_{\Lt_l}= \frac{1}{1+\ell\left(\theta,\theta;\frac{2}{\alpha}\right)\sum_{m=1}^{K-1}\hat{\vartheta}_m},\label{Eqn:CrossRATCovProbRATL}\\
\underline{\hat{\mathsf{P}}_{\Lt_u}}&=\lim_{\mu\rightarrow\infty}\hat{\mathsf{P}}_{\Lt_u}=\sum_{k=1}^{K-1}\mathbb{E}_{\frac{G'_k}{G_k}}\left\{\left(1+ \sum_{m=1}^{K} p_m\hat{\vartheta}_m\mathbb{E}_{\frac{G'_m}{G_m}}\left[\ell\left(\hat{\Theta}_{k,m},\hat{\Theta}_{k,m};\frac{2}{\alpha}\right)\right]\right)^{-1}\right\}\vartheta_k,\label{Eqn:CrossRATCovProbRATLU}
\end{align}
\begin{align}
\underline{\hat{\mathsf{P}}_{\Ut}}&=\lim_{\mu,\tau^{-1}_k\rightarrow\infty}\hat{\mathsf{P}}_{\Ut}=\left\{1+p_K\hat{\vartheta}_K\ell\left(\theta,\theta;\frac{2}{\alpha}\right)+\sum_{m=1}^{K-1}p_m\hat{\vartheta}_m\mathbb{E}\left[\ell\left(\frac{\theta}{G}_m,\frac{\theta}{G}_m;\frac{2}{\alpha}\right)\right]\right\}^{-1},\label{Eqn:CrossRATCovProbRATU}
\end{align}
respectively, and they are apparently greater than those limits in \eqref{Eqn:RATLCovProbLowLimit}, \eqref{Eqn:LowLimitRatLuCovProb} and \eqref{Eqn:LowLimitRatUCovProb2}. The lowest limit on $\hat{\mathsf{P}}_{\cov}$ is readily found as $\underline{\hat{\mathsf{P}}_{\cov}}=\sum_{k=1}^{K-1}\vartheta_k\underline{\hat{\mathsf{P}}_{\Lt_l}}+\vartheta_K\underline{\hat{\mathsf{P}}_{\Ut}}$.  These lowest limits can be used to expressed the lowest limit on the coexisting network capacity defined in the following subsection for the crossing-RAT user association scenario. 
 
\subsection{Mean Spectrum Efficiency, Coexisting Network Capacity and Their Limits}
According to Definition \ref{Defn:CoeNetCapa}, the coexisting network capacity for the crossing-RAT scenario can be also defined as
\begin{align}
\hat{\mathsf{C}}_{\cov}=\sum_{k=1}^{K-1}\lambda_k (1-\hat{\nu}_k) \hat{\mathsf{P}}_{\Lt}\hat{\mathsf{C}}_{\Lt} +\lambda_K(1-\hat{\nu}_K) \hat{\mathsf{P}}_{\Ut}\hat{\mathsf{C}}_{\Ut},
\end{align}
where $\hat{\mathsf{C}}_{\Ut}\defn \hat{\rho}_Kp_K\mathbb{E}\left[\log_2(1+\hat{\gamma}_{\Ut})\right]$ is the mean spectrum efficiency of an RAT-$\Ut$ AP and $\hat{\mathsf{C}}_{\Lt}$ is the total mean spectrum efficiency of an RAT-$\Lt$ AP given by
\begin{align}
\hat{\mathsf{C}}_{\Lt}\defn \mathbb{E}\left[\log_2(1+\hat{\gamma}_{\Lt_l})+\left(\sum_{k=1}^{K-1}\hat{\rho}_kp_k\vartheta_k\right)\log_2(1+\hat{\gamma}_{\Lt_u})\right]. 
\end{align} 
The explicit results of  the lower bounds on $\hat{\mathsf{C}}_{\Lt}$ and $\hat{\mathsf{C}}_{\Ut}$ can be derived by the lower bounds on $\hat{\mathsf{P}}_{\Lt}$ and $\hat{\mathsf{P}}_{\Ut}$, respectively, as the integral method shown in the proof of Corollary \ref{Thm:LowerBoundMeanRate}. They are used to characterize the lower bound on $\hat{\mathsf{C}}_{\cov}$, as shown in the following corollary.

\begin{corollary}\label{Thm:CrossRATCapacity}
According to the coverage results in Theorem \ref{Thm:CrossRATCoeCoverage}, the lower bound on the coexisting network capacity can be shown as
\begin{align}
\hat{\mathsf{C}}_{\cov}\geq & \sum_{k=1}^{K-1} \lambda_k(1-\hat{\nu}_k) \hat{\mathsf{P}}_{\Lt}(\theta)\left(\int_{0}^{\infty}\frac{\hat{\mathsf{P}}_{\Lt_l}(\theta)+\sum_{k=1}^{K-1}(\hat{\rho}_kp_k\vartheta_k)\hat{\mathsf{P}}_{\Lt_u}(\theta)}{(\ln 2)(1+\theta)}\dif \theta\right) \nonumber \\
&+\lambda_K(1-\hat{\nu}_K)\hat{\rho}_Kp_K\hat{\mathsf{P}}_{\Ut}(\theta) \int_{0}^{\infty}\frac{\hat{\mathsf{P}}_{\Ut}(\theta)}{(\ln 2)(1+\theta)}\dif \theta,
\end{align} 
where $\hat{\mathsf{P}}_{\Lt}(\theta)$ and $\hat{\mathsf{P}}_{\Ut}(\theta)$ are given in \eqref{Eqn:CrossRATCovProbRATLu} and \eqref{Eqn:CrossRATCovRATU}, respectively.
\end{corollary}
\begin{IEEEproof}
The proof is omitted here since it is similar to the proof of Corollary. \ref{Thm:LowerBoundMeanRate}. 
\end{IEEEproof}
\noindent The coexisting network capacity in the scenario of crossing-RAT user association is surely higher than that in the scenario of noncrossing-RAT user association since the link coverages achieved by crossing-RAT user association are higher than those achieved by noncrossing-RAT user association. In other words, the network capacity can increase if users can do crossing-RAT user association. From the viewpoint of user's link capacity, however, the link capacity of users may not increase because cross-RAT user association does not reflect the benefit of the additional capacity incrementation when an RAT-$\Lt$ AP can access the RAT-$\Ut$ channel. Therefore, when a user associates with an RAT-$\Ut$ AP its link capacity may not be higher than that if it associated with an RAT-$\Lt$ AP even though the RAT-$\Ut$ AP can provide the highest SIR to it.  

\section{Numerical Simulation for Coexisting LTE-U and WiFi Networks}\label{Sec:Simulation}

\begin{table}[!t]
\centering
\caption{Network Parameters for Simulation}\label{Tab:SimPara}
\begin{tabular}{|c|c|c|c|c|}
\hline Parameter $\setminus$ AP Type (Tier \#)& Macrocell (1) & Picocell (2) & Femtocell (3) & WiFi (4)\\ \hline
Power $P_k$ (W) & 40 & 1  & 0.5 & 0.2 \\ \hline
Intensity $\lambda_k$ (APs/$m^2$) & $1\times 10^{-6}$ & $10\lambda_1$ & $50\lambda_1$ & $100\lambda_1$   \\ \hline
Maximum Backoff Time $\tau_k$ & $\infty$ &\multicolumn{2}{c|}{2}& 1 \\ \hline
Sensing Area $\mathcal{S}_k$ ($m^2$) & N/A &\multicolumn{3}{c|}{900$\pi$}  \\ \hline
CSMA Threshold $\Delta$ & N/A & \multicolumn{3}{c|}{$4.481$}  \\ \hline 
SIR Threshold $\theta$ & \multicolumn{4}{c|}{0.5} \\ \hline  
$G_{k,i}$ & \multicolumn{4}{c|}{$\sim \ln\mathcal{N}(0,3\text{dB})$} \\ \hline
Pathloss Exponent $\alpha$ &\multicolumn{4}{c|}{4}\\ \hline 
\end{tabular} 
\end{table}

In this section, we apply our previous modeling and analysis framework to the scenario in which LTE small cell BSs and WiFi APs coexist to access the unlicensed frequency band. Here our objective is to numerically evaluate how link coverage and capacity of the WiFi APs are affected by LTE small cell BSs. We consider there are four tiers in the HetNet -- the first three tiers consisting of the macro BSs, picocells and femtocells belonging to the LTE cellular subnetwork and the fourth tier consisting of the APs belonging to the WiFi subnetwork. Namely, LTE small cell BSs primarily use the licensed band channel (LTE-$\Lt$ channel) and opportunistically access their unlicensed band (LTE-$\Ut$) channel  whereas WiFi APs only access their unlicensed band  (WiFi-$\Ut$) channel by opportunistic CSMA/CA. Assuming the MMPA scheme is adopted, macro BSs do not access the channel in the unlicensed band, and the random backoff time  of the picocells and femtocells with opportunistic CSMA/CA is uniformly-distributed so that their channel access probability can be found by \eqref{Eqn:TransProbUniDis}. The network parameters for simulation are listed in Table \ref{Tab:SimPara}. 

\begin{figure}[t!]
	\centering
	\includegraphics[width=5.5in,height=2.5in]{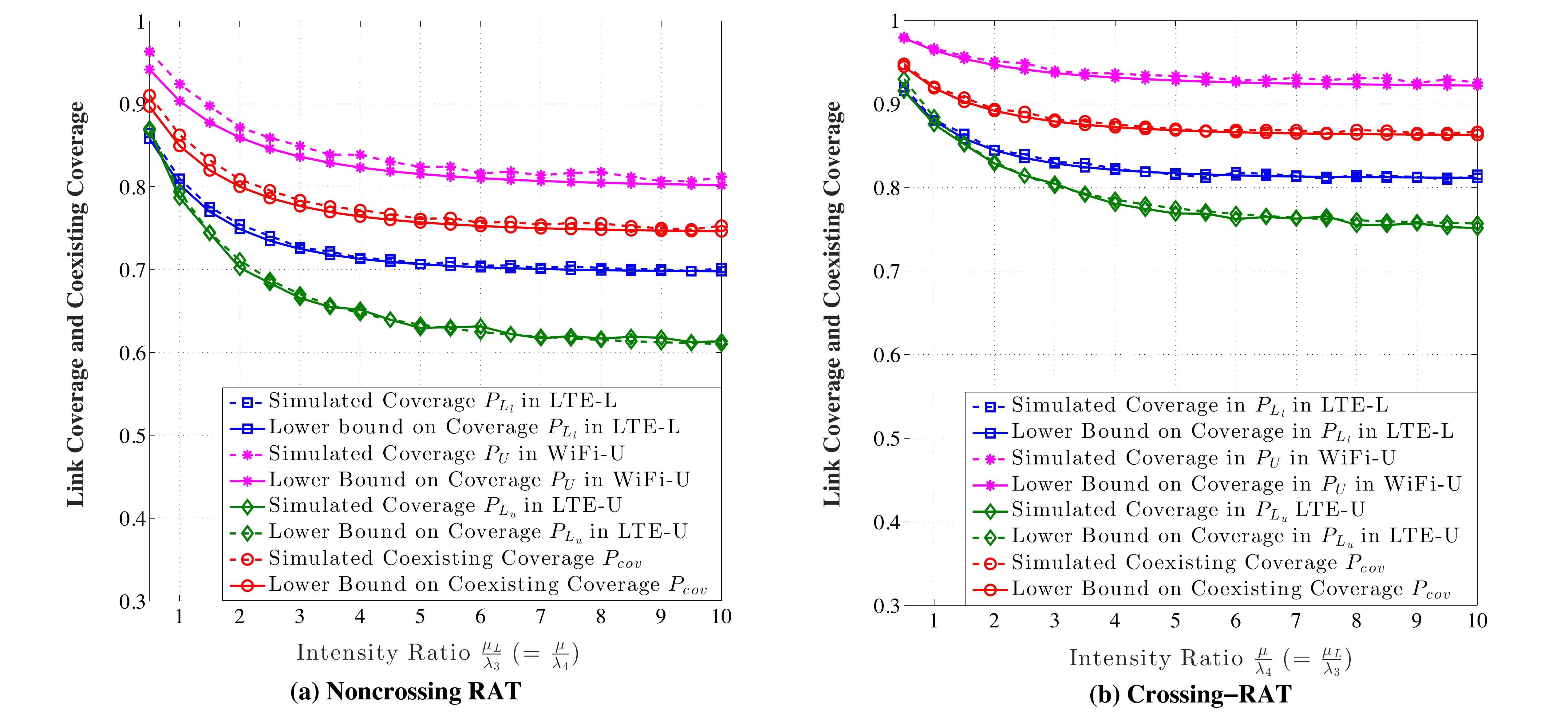}
	\caption{(a) Link coverage and coexisting coverage for noncrossing-RAT user association and user intensity $\mu_{\Lt}=\mu_{\Ut}$, (b) Coverage and coexisting coverage for crossing-RAT user association and user intensity $\mu=2\mu_L$. Note that the horizontal axises of the two sub-figures have the same scale since $\frac{\mu_L}{\lambda_3}=\frac{\mu}{\lambda_4}$.}
	\label{Fig:CoeCovProb}
\end{figure}

\begin{figure}[t!]
	\centering
	\includegraphics[width=5.5in,height=2.5in]{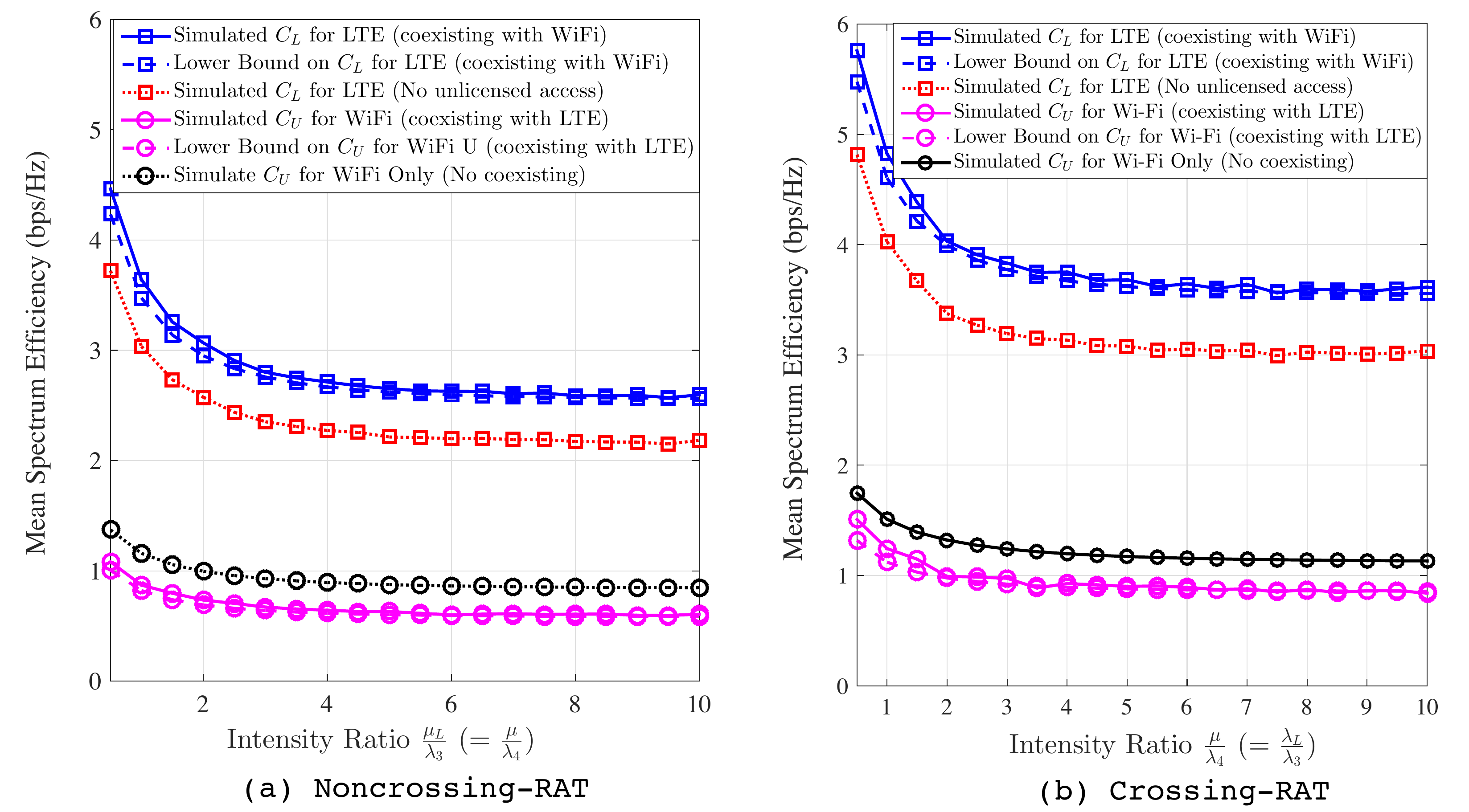}
	\caption{(a) Mean Spectrum Efficiency for noncrossing-RAT user association and user intensity $\mu_{\Lt}=\mu_{\Ut}=\frac{1}{2}\mu$, (b) Mean Spectrum Efficiency for crossing-RAT user association and user intensity $\mu=2\mu_{\Lt}$. Note that the horizontal axises of the two sub-figures have the same scale since $\frac{\mu_L}{\lambda_3}=\frac{\mu}{\lambda_4}$.}
	\label{Fig:MeanSpectrumEff}
\end{figure}

\begin{figure}[t!]
	\centering
	\includegraphics[width=5.5in,height=2.5in]{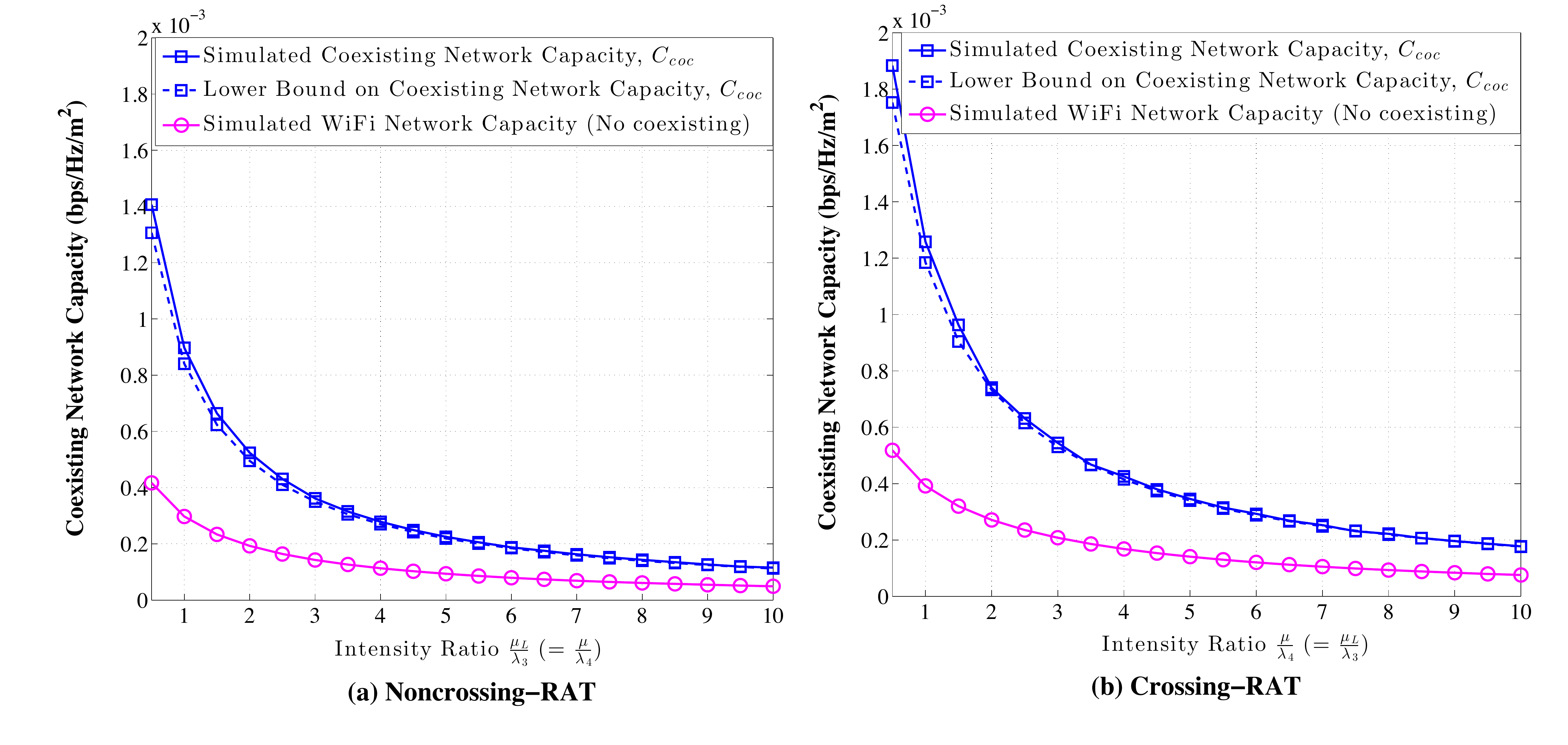}
	\caption{(a) Coexisting network capacity for noncrossing-RAT user association and user intensity $\mu_{\Lt}=\mu_{\Ut}$, (b) Coexisting network capacity for crossing-RAT user association and user intensity $\mu=2\mu_{\Lt}$. Note that the horizontal axises of the two sub-figures have the same scale since $\frac{\mu_L}{\lambda_3}=\frac{\mu}{\lambda_4}$.}
	\label{Fig:CoexistingNetworkCapacity}
\end{figure}

The simulation results for the link coverages are shown in Fig. \ref{Fig:CoeCovProb}. As we can see in the figure, all simulated results are fairly close to their corresponding lower bounds, which validates that the derived lower bounds on the link coverage and coexisting coverage in Theorems \ref{Thm:CoexistCoverage} and \ref{Thm:CrossRATCoeCoverage} are very tight and accurate. Hence, we can realize that using PPPs to approximate the non-PPP BSs induced by user association still can lead to a very accurate result in general. Surprisingly, the lower bounds on the link coverages derived by using PPPs to approximate the MHPPs of the WiFi APs, picocell and femtocell BSs in the unlicensed band are still very accurate as well. This is because the channel-aware opportunistic scheduling helps CSMA/CA alleviate the location correlations between the LTE BSs and the WiFi APs and makes the MHPPs become more like PPPs. In order to make WiFi APs not be affected too much while the LTE small cell BSs are accessing the unlicensed band channel, we let WiFi APs have a short backoff time range so that they have a higher link coverage and chance to access the unlicensed band channel than the the LTE BSs as shown in Fig. \ref{Fig:CoeCovProb}. Thus, CSMA/CA with random backoff time is principally similar to the Listen-before-Talk (LBT) with Carrier Sensing Adaptive Transmission (CSAT) and Licensed-Assisted Access (LAA) protocols proposed in the LTE-U\cite{HZXCWGSW15}\cite{RZMWLXCZZXSLLX15}. Thus, the numerical results in this section are a good reference for evaluating the network-wise performance of CSAT and LAA. All the link and coexisting coverages in Fig. \ref{Fig:CoeCovProb}(b) are much better than those in Fig. \ref{Fig:CoeCovProb}(a), as expected, since crossing-RAT makes users exploit more ``multi-AP diversity'' while doing user association. Also, all coverages decrease along the user intensity and eventually converge to their lowest limits since the void probabilities reduce to zero as the user intensity goes to infinity. 

The simulation results for the mean spectrum efficiencies of LTE users and WiFi users are shown in Fig. \ref{Fig:MeanSpectrumEff}, i.e., $\mathsf{C}_{\Lt}$ defined in \eqref{Eqn:DefnMeanCapRATL} and $\mathsf{C}_{\Ut}$ defined in \eqref{Eqn:DefnMeanCapRATU}, and we can see the bower bounds on $\mathsf{C}_{\Lt}$ and $\mathsf{C}_{\Ut}$ given in \eqref{Eqn:RATLMeanRate} and \eqref{Eqn:RATUMeanRate} are pretty tight and close to their corresponding simulation results. Most importantly, we observe that the sum of $\mathsf{C}_{\Lt}$ and $\mathsf{C}_{\Ut}$ is much higher than the mean spectrum efficiency of the WiFi APs without coexisting with LTE small cell BSs (i.e., the ``WiFi only'' result). This certainly implies that there potentially exists a considerable link capacity gain if LTE and WiFi can coexist well. Fig. \ref{Fig:CoexistingNetworkCapacity} shows the simulation results of the coexisting network capacities that are almost equal to their derived lower bounds, as expected. The coexisting network capacity in each user association scenario is much higher than the network capacity of the WiFi  subnetwork without coexisting with the LTE small cell subnetwork. Accordingly, making different RAT networks coexist favorably is able to bring a notable improvement in the overall network capacity.

\section{Conclusion}
In this paper, a modeling and analysis framework is proposed for a multi-RAT HetNet with two scenarios of crossing-RAT and noncrossing-RAT user associations. For each user association scenario, we first derive the void cell probability and the RAT-$\Ut$ channel access probability of the opportunistic CSMA/CA protocol for the APs in each tier. To evaluate the coexistence interplay between the APs, the coexisting coverage and network capacity are proposed and their tight lower bounds and lowest limits are found in closed-form. Our salient findings are to show that the coverage and capacity are both significantly improved based on our more realistic modeling framework for HetNets, the opportunistic CSMA/CA protocol induces much less interference and location correlation between the transmitting APs so that the link and coexisting coverages and capacities can be accurately estimated by their derived lower bounds, and crossing-RAT user association can achieve higher coverages and capacities than noncrossing-RAT. Numerical simulations verify that the transmission performance of coexisting LTE-U and WiFi APs can be well-characterized by the proposed multi-RAT modeling and analyzing approaches and all the derived lower bounds on the coverages and capacities are very tight and accurate.

\appendices
\section{User Association Statistics}\label{App:LemmaNonCrossUserAss}
\begin{lemma}\label{Lem:StaDisUserAss}
Suppose the fraction moments of all $W_{k,i}$'s in \eqref{Eqn:NonCrossUserAssSch} exist, i.e., $\mathbb{E}\left[W_k^a\right]<\infty$ for all $a\in (0,1)$ and $k\in\mathcal{K}$. For the noncrossing-RAT user association scenario, if the associated AP  $X^*_{\Rt}$ in \eqref{Eqn:NonCrossUserAssSch} uses RAT-$\Lt$, the cumulative distribution function (CDF) of the weighted distance $\|\tilde{X}^*_{\Lt}\|\defn(W^*_{\Lt})^{-\frac{1}{\alpha}}\|X^*_{\Lt}\|$ can be shown as
\begin{align}
F_{\|\tilde{X}^*_{\Lt}\|}(x)=1-e^{-\pi x^2\sum_{k=1}^{K-1}\lambda_k\mathbb{E}\left[W_{k}^{\frac{2}{\alpha}}\right]}\label{Eqn:CDF_AssAP-RATL}
\end{align}
and $\tilde{X}^*_{\Lt}$ can be viewed as the node in a homogeneous PPP of intensity $\sum_{k=1}^{K-1}\lambda_k\mathbb{E}\left[W_k^{\frac{2}{\alpha}}\right]$ nearest to the origin. Moreover, the distribution of $\|\tilde{X}^*_k\|$ given that $X^*_{\Lt}$ is from the $k$th tier is the same as that of $\|\tilde{X}^*_{\Lt}\|$, i.e., $F_{\|\tilde{X}^*_k\|}(x)=F_{\|\tilde{X}^*_{\Lt}\|}(x)$. Also, if $X^*_{\Rt}$ in \eqref{Eqn:NonCrossUserAssSch} adopts RAT-$\Ut$  the CDF of the weighted distance $\|\tilde{X}^*_{\Ut}\|\defn(W^*_{\Ut})^{-\frac{1}{\alpha}}\|X^*_{\Ut}\|$ can be directly found by \eqref{Eqn:CDF_AssAP-RATL} as
\begin{align}
F_{\|\tilde{X}^*_{\Ut}\|}(x)= 1-e^{-\pi x^2\lambda_K\mathbb{E}\left[W^{\frac{2}{\alpha}}_K\right]}. \label{Eqn:CDF_AssAP-RATU}
\end{align}
For the crossing-RAT user association scenario, if the associated AP $X^*$ is in \eqref{Eqn:CrossUserAss} and the distance from it to the origin is given in \eqref{Eqn:CrossUserAssDis}, $\tilde{X}^*$ can be viewed the node in a homogeneous PPP of intensity $\sum_{k=1}^{K}\lambda_k\mathbb{E}\left[W^{\frac{2}{\alpha}}_k\right]$ nearest to the origin and the CDF of the distance from it to the origin is
\begin{align}
F_{\|\tilde{X}^*\|}(x) = 1-e^{-\pi x^2\sum_{k=1}^{K}\lambda_k\mathbb{E}\left[W_{k}^{\frac{2}{\alpha}}\right]},\label{Eqn:CDF_CrossRATAssAP}
\end{align}
which is also equal to the CDF of the weighted distance $\|\tilde{X}^*_k\|\defn(W^*)^{-\frac{1}{\alpha}}\|X^*\|$ for given $X^*\in\Phi_k$, i.e., $F_{\|\tilde{X}^*\|}(x) =F_{\|\tilde{X}^*_k\|}(x) $.
\end{lemma}
\begin{IEEEproof} 
According to \eqref{Eqn:NonCrossUserAssSchDist}, the CDF of the weighted distance $\|\tilde{X}^*_{\Lt}\|$ can be shown as follows
\begin{align*}
F_{\|\tilde{X}^*_{\Lt}\|}(x)& \defn\mathbb{P}\left[(W^*_{\Lt})^{-\frac{1}{\alpha}}\|X^*_{\Lt}\|\leq x\right]=1-\mathbb{P}\left[\sup_{X_{k,i}\in\bigcup_{k=1}^{K-1}\Phi_k} W_{k,i}\|X_{k,i}\|^{-\alpha}\leq x^{-\alpha}\right]\\
&=1-\mathbb{E}\left\{\prod_{X_{k,i}\in\bigcup_{k=1}^{K-1}\Phi_k}\mathbb{P}\left[ \frac{W_{k,i}}{\|X_{k,i}\|^{\alpha}}\leq x^{-\alpha}\right]\right\}\stackrel{(\star)}{=} 
1-e^{-2\pi\sum_{k=1}^{K-1}\lambda_k\int_{0}^{\infty} \mathbb{P}\left[W^{\frac{1}{\alpha}}_kx\geq r\right]r\dif r},
\end{align*}
where $(\star)$ follows from the probability generating functional (PGF) of a homogeneous PPP\cite{DSWKJM96}\cite{MHRKG09}. Since $2\int_{0}^{\infty} \mathbb{P}\left[W^{\frac{1}{\alpha}}_kx\geq r\right]r\dif r=x^2\mathbb{E}\left[W^{\frac{2}{\alpha}}\right]$,  $F_{\|\tilde{X}^*_{\Lt}\|}(x)$ in  \eqref{Eqn:CDF_AssAP-RATL} is obtained. For given $X^*_{\Lt}\in\Phi_k$, we have $F_{\|\tilde{X}^*_k\|}(x)=\mathbb{P}\left[(W^*_{\Lt})^{-\frac{1}{\alpha}}\|X^*_{\Lt}\|\leq x|X^*_{\Lt}\in\Phi_k\right]=\frac{1}{\vartheta_k}\mathbb{P}\left[W^*_{\Lt}\|X^*_{\Lt}\|^{-\alpha}\geq x^{-\alpha},X^*_{\Lt}\in\Phi_k\right]$. By letting $Z_{-k}=\sup_{X_{m,i}\in\bigcup_{m\in\mathcal{K}\setminus k}\Phi_m} W_{m,i}\|X_{m,i}\|^{-\alpha}$ and $Z_k=W^*_{\Lt}\|X^*_{\Lt}\|^{-\alpha}=\sup_{X_{k,i}\in\Phi_k}W_{k,i}\|X_{k,i}\|^{-\alpha}$ for $X^*_{\Lt}\in\Phi_k$, it follows that $F_{\|\tilde{X}^*_k\|}(x) = \frac{1}{\vartheta_k} \mathbb{P}\left[ Z_k\geq \max\left\{x^{-\alpha}, Z_{-k}\right\}\right]=\frac{1}{\vartheta_k}\mathbb{P}\left[ Z^{-\frac{1}{\alpha}}_k\leq \min\left\{x, Z^{-\frac{1}{\alpha}}_{-k}\right\}\right]$ and $F_{\|\tilde{X}^*_k\|}(x) = \frac{1}{\vartheta_k}\left( 1-\int_{0}^{x}e^{-\pi\lambda_kz^2}\dif F_{Z_{-k}}(z)-e^{-\pi\lambda_kx^2}\int_{x}^{\infty} \dif F_{Z_{-k}}(z)\right)$,
which equals to \eqref{Eqn:CDF_AssAP-RATL} because $F_{Z_{-k}}(z)=1-e^{-\pi z^2\sum_{m\in\mathcal{K}\setminus k}\lambda_m\mathbb{E}\left[G^{\frac{2}{\alpha}}_m\right]}$. Using the similar steps of showing \eqref{Eqn:CDF_AssAP-RATL} and \eqref{Eqn:CDF_AssAP-RATU} in above, the result in \eqref{Eqn:CDF_CrossRATAssAP} can be obtained.  
\end{IEEEproof}



\section{Proof of Theorem \ref{Thm:CoexistCoverage}}\label{App:ProofCoexistCoverage}
First, we show the lower bound on the RAT-$\Lt$ link coverage in \eqref{Eqn:RatLCovProb}. According to \eqref{Eqn:NonCrossUserAssSchDist} for the MMPA scheme and $\gamma_{\Lt}$ in \eqref{Eqn:RATL-SIR}, the RAT-$\Lt$ link coverage can be equivalently expressed as
\begin{align*}
\mathsf{P}_{\Lt_l}= \sum_{k=1}^{K-1}\mathbb{P}\left[\frac{H_kP^*_{\Lt}(G^*_{\Lt})^{-1}}{I_{\Lt_l}\|X^*_{\Lt}\|^{\alpha}}\geq\theta\right]\mathbb{P}\left[X^*_{\Lt}\in\Phi_k\right]\stackrel{(a)}{=}\mathbb{P}\left[\frac{H^*_{\Lt}P^*_{\Lt}(G^*_{\Lt})^{-1}}{I_{\Lt_l}\|X^*_{\Lt}\|^{\alpha}}\geq\theta\right],
\end{align*}
where $H^*_{\Lt}$ is an exponential random variable with unit mean and variance, and $I_{\Lt_l}$ is defined in \eqref{Eqn:RATL-SIR} and $(a)$ follows from the fact that all fading channel gains are i.i.d. and the mean signal power that is the maximum mean received power among all mean received powers from all RAT-$\Lt$ APs does not depend on any specific tier index. According to Lemma \ref{Lem:StaDisUserAss} in Appendix \ref{App:LemmaNonCrossUserAss}, we can have
\begin{align*}
\mathsf{P}_{\Lt_l}=\mathbb{P}\left[\frac{H^*_{\Lt}}{\tilde{I}_{\Lt_l}\|\tilde{X}^*_{\Lt}\|^{\alpha}}\geq\theta\right]=\mathbb{E}\left[\exp\left(-\theta \tilde{I}_{\Lt_l}\|\tilde{X}^*_{\Lt}\|^{\alpha}\right)\right],
\end{align*}
where $\tilde{I}_{\Lt_l}=\sum_{m,i:\tilde{X}_{m,i}\in\tilde{\Psi}_{\Lt}\setminus\tilde{X}^*_{\Lt}} V_{m,i}H_{m,i}\|\tilde{X}_{m,i}\|^{-\alpha}$, $\tilde{\Psi}_{\Lt}\defn \bigcup_{m=1}^{K-1}\tilde{\Phi}_m$ where $\tilde{\Phi}_m\defn\{\tilde{X}_{m,i}\in\mathbb{R}^2: \tilde{X}_{m,i}=(P_mG^{-1}_{m,i})^{-\frac{1}{\alpha}}X_{m,i}, X_{m,i}\in \Phi_m\}$ and it is a PPP of intensity $\sum_{m=1}^{K-1}\lambda_m P^{\frac{2}{\alpha}}_m\mathbb{E}\left[G^{-\frac{2}{\alpha}}_m\right]$, $\tilde{X}^*_{\Lt}\defn P^{-\frac{1}{\alpha}}_k(G^*_k)^{\frac{1}{\alpha}} X^*_{\Lt}$ and it can be viewed as the AP in $\tilde{\Psi}_{\Lt}$ nearest to the origin. Since $V_{m,i}$'s may not be independent due to the location correlations between APs induced by user association \cite{CHLLCW1502,CHLLCW16}, the closed-form result of $\mathsf{P}_{\Lt}$ is unable to be obtained. However, its lower bound can be derived by assuming all $V_{m,i}$'s are independent and such an assumption makes the non-void APs become a thinning PPP that generates to a larger interference power since they are able to be arbitrarily close to the typical user while the original location-correlated non-void APs are not. Thus, using the proof techniques of Proposition 2 in \cite{CHLLCW16}, the lower bound can be derived as shown in the following:
\begin{align*}
\mathsf{P}_{\Lt_l}&\geq\mathbb{E}\left[e^{-\pi\sum_{k=1}^{K-1}\lambda_k(1-\nu_k)P^{\frac{2}{\alpha}}_k\mathbb{E}\left[G^{-\frac{2}{\alpha}}_k\right]\|\tilde{X}^*_{\Lt}\|^2}\right]\stackrel{(b)}{=}2\pi\left(\sum_{k=1}^{K-1}\lambda_kP^{\frac{2}{\alpha}}_k\mathbb{E}\left[G^{-\frac{2}{\alpha}}_k\right]\right)\times\\
&\int_{0}^{\infty} e^{-\pi\sum_{k=1}^{K-1}\lambda_kP^{\frac{2}{\alpha}}_k\mathbb{E}\left[G^{-\frac{2}{\alpha}}_k\right]\left(\ell\left(\theta,\theta;\frac{2}{\alpha}\right)\sum_{k=1}^{K-1}(1-\nu_k)\vartheta_k+1\right)x^2} x\dif x=\int_{0}^{\infty} e^{-\left(1+\ell\left(\theta,\theta;\frac{2}{\alpha}\right)\sum_{k=1}^{K-1}(1-\nu_k)\vartheta_k\right)}y\dif y
\end{align*}
where $(b)$ follows from the result in Lemma \ref{Lem:StaDisUserAss} in Appendix \ref{App:LemmaNonCrossUserAss} that indicates the distribution of $\|\tilde{X}^*_{\Lt}\|$ is the same no matter which tier $\tilde{X}^*_{\Lt}$ belongs to. Then carrying out the  last integral yields the result in \eqref{Eqn:RatLCovProb}. The link coverage of the RAT-$\Lt$ users in the RAT-$\Ut$ channel is $\mathbb{P}\left[\gamma_{\Lt_u}\geq \theta\right]$ where $\gamma_{\Lt_u}$ is given in \eqref{Eqn:RATL-SIR}, and its identity can be shown by using Lemma \ref{Lem:StaDisUserAss} as
\begin{align}
\mathsf{P}_{\Lt_u}=\sum_{k=1}^{K-1}\mathbb{E}\left[e^{-\theta G'_kG^{-1}_kG_kP^{-1}_k\|X^*_{\Lt}\|^{\alpha}I_{\Lt_u}}\big|X^*_{\Lt}\in\Phi_k\right]\vartheta_k=\sum_{k=1}^{K-1}\mathbb{E}\left[e^{-\theta G'_kG^{-1}_k\|\tilde{X}_{\Lt}\|^{\alpha}\tilde{I}'_{\Lt_u}}\right]\vartheta_k,\label{Eqn:ProofCovProbRATLu}
\end{align}
where $\tilde{I}'_{\Lt_u}=\sum_{m,i:\tilde{X}_{m,i}\in\tilde{\Psi}_{\Ut}\setminus\tilde{X}^*_{\Lt}}V_{m,i}H_{m,i}G'_{m,i}G^{-1}_{m,i}\|\tilde{X}_{m,i}\|^{-\alpha}$ with $\tilde{\Psi}_{\Ut}\defn\{\tilde{X}_{m,i}\in\mathbb{R}^2: \tilde{X}_{m,i}=P^{-\frac{1}{\alpha}}_{m,i}G'^{\frac{1}{\alpha}}_{m,i}X_{m,i}, X_{m,i}\in\Psi_{\Ut}\}$ is not a homogeneous PPP but a Mat\'{e}rn hard-core point process (MHPP) of intensity $\sum_{m=1}^K \lambda^{\dagger}_m$ where $\lambda^{\dagger}_m\defn \rho_mp_m\lambda_mP^{\frac{2}{\alpha}}_m\mathbb{E}\left[G^{-\frac{2}{\alpha}}_m\right]$ due to the opportunistic CSMA/CA protocol\cite{FBBBL10,XDCHLLCWXZ16}. Modeling $\tilde{\Psi}_{\Lt_u}$ as a PPP yields the lower bound on $\mathbb{E}\left[e^{-\theta G_k\|\tilde{X}^*_{\Lt}\|^{\alpha}\tilde{I}'_{\Lt_u}}\right]$ since the APs in a PPP are able to be arbitrarily close to the typical user so that they generates a larger interference than the APs in an MHPP that are not allowed to be arbitrarily close to the typical user.  Hence, letting $\Omega_{\frac{k}{k'},\frac{m}{m'}}(\theta)\defn \mathbb{E}_{\frac{G'_m}{G_m}}\left[\int_{1}^{\infty} \frac{\dif y}{1+\left(\frac{G_kG_m}{\theta G'_k  G'_m}\right)y^{\frac{\alpha}{2}}}\right]$ yields
\begin{align*}
\mathbb{E}\left[e^{-\theta G'_kG^{-1}_k\|\tilde{X}^*_{\Lt}\|^{\alpha}\tilde{I}'_{\Lt_u}}\right] 
\stackrel{(c)}{\geq}& \mathbb{E}_{\frac{G'_k}{G_k}}\left[e^{- \sum_{m=1}^{K-1}\pi(1-\nu_m)\lambda^{\dagger}_m\|\tilde{X}^*_{\Lt}\|^2\Omega_{\frac{k}{k'},\frac{m}{m'}}(\theta)}\right]\mathbb{E}_{\frac{G'_k}{G_k}}\left[e^{- \pi(1-\nu_K)\lambda^{\dagger}_K\|\tilde{X}^*_{\Lt}\|^2 \Omega_{\frac{K}{K'},\frac{k}{k'}}(\theta)}\right]\\
\stackrel{(d)}{\geq}& \mathbb{E}_{\frac{G'_k}{G_k}}\left\{\left(1+\sum_{m=1}^{K}(1-\nu_m)\rho_mp_m\vartheta_m\mathbb{E}_{G_m}\left[\ell\left(\frac{G'_mG'_k}{G_mG_k}\theta ,\Theta_{k,m};\frac{2}{\alpha}\right)\right]\right)^{-1}\right\},
\end{align*}
where $(c)$ is due to modeling the resulting transmitting APs as $K-1$ independent thinning PPPs that generate larger interference and $\tilde{X}^*_{\Lt}$ is from the first $K-1$ tiers, and $(d)$ is obtained by averaging over $\|\tilde{X}^*_{\Lt}\|^2$. Then substituting this inequality result into \eqref{Eqn:ProofCovProbRATLu} leads to \eqref{Eqn:CovProbRATLU}.

Now we show how to find the lower bound on $\mathsf{P}_{\Ut}$. The explicit expression of $\mathsf{P}_{\Ut}$ is given by
\begin{align*}
\mathsf{P}_{\Ut} = \mathbb{P}\left[H^*_KP_K(G^*_K)^{-1}\geq\theta I_{\Ut}\|X^*_{\Ut}\|^{\alpha}\right]=\mathbb{E}\left[\exp\left(-\theta(\tilde{I}_{\Lt_u}+\tilde{I}_{\Ut_u})\|\tilde{X}^*_{\Ut}\|^{\alpha}\right)\right],
\end{align*}
where $\tilde{I}_{\Lt_u}\defn\sum_{\tilde{X}_{m,i}\in\tilde{\Psi}_{\Ut}\setminus\tilde{\Phi}_K}V_{m,i}\|\tilde{X}_{m,i}\|^{-\alpha}$, $\tilde{X}^*_{\Ut}\defn P^{-\frac{1}{\alpha}}_K(G^*_K)^{\frac{1}{\alpha}}X^*_{\Ut}$ and it is the nearest point in $\tilde{\Phi}_K$ to the origin, and $\tilde{I}_{\Ut_u}\defn\sum_{\tilde{X}_{m,i}\in\tilde{\Phi}_K\setminus \tilde{X}^*_{\Ut}}V_{m,i}\|\tilde{X}_{m,i}\|^{-\alpha}$. Since $\tilde{\Psi}_{\Ut}$ is an MHPP and all $V_{m,i}$'s are not completely independent, the closed-form expression of $\mathsf{P}_{\Ut}$ is essentially unable to be found so that its lower bound can be found by assuming $\tilde{\Psi}_{\Ut}$ is a PPP and $V_{m,i}$'s are all independent. Since $\tilde{I}_{\Lt_u}$ and $\tilde{I}_{\Ut_u}$ are independent, it follows that
\begin{align}
\mathsf{P}_{\Ut}\geq & \mathbb{E}_{\|\tilde{X}^*_{\Ut}\|}\left\{\mathbb{E}\left[e^{-\theta(\tilde{I}_{\Lt_u}+\tilde{I}_{\Ut_u})\|\tilde{X}^*_{\Ut}\|^{\alpha}}\bigg|\|X^*_{\Ut}\|\right] \right\}
\stackrel{(e)}{=} \mathbb{E}_{\|\tilde{X}^*_{\Ut}\|}\left\{e^{-\pi\|\tilde{X}^*_{\Ut}\|^2\sum_{k=1}^{K}\ell\left(\theta,\theta_k;\frac{2}{\alpha}\right)\lambda^{\dagger}_k(1-\nu_k)}\right\},\label{Eqn:ProofLowBoundRATUCovProb}
\end{align}
where  $(e)$ follows from the results of the Laplace transforms of $\tilde{I}_{\Lt_u}$ and $\tilde{I}_{\Ut_u}$ for a given $\|\tilde{X}^*_{\Ut}\|$ and note that $\|\tilde{X}^*_{\Ut}\|^{-\alpha}$ is independent from $\tilde{I}_{\Lt_u}$ and is the maximum term in $\tilde{I}_{\Ut_u}$\cite{MHRKG09,CHLLCW16}. Furthermore, the lower bound in \eqref{Eqn:RatUCovProb} can be found by averaging the lower bound in \eqref{Eqn:ProofLowBoundRATUCovProb}  over $\|\tilde{X}^*_{\Ut}\|$ since the pdf of $\|\tilde{X}^*_{\Ut}\|$ is $f_{\|\tilde{X}^*_{\Ut}\|}(x)=2\pi\lambda_KP^{\frac{2}{\alpha}}_K\mathbb{E}\left[G^{-\frac{2}{\alpha}}_K\right] x e^{-\pi\lambda_KP^{\frac{2}{\alpha}}_K\mathbb{E}[G^{-\frac{2}{\alpha}}_K] x^2}$. The lower bound on $\mathsf{P}_{\cov}$ can be acquired by the two lower bounds on $\mathsf{P}_{\Lt}$ and $\mathsf{P}_{\Ut}$.

\bibliographystyle{ieeetran}
\bibliography{IEEEabrv,Ref_MultiRATHetNets}

\end{document}